# Decentralized autonomous organization and blockchain-based incentivization framework for community-based facilities management

Reachsak Ly, Ph.D.[1] Alireza Shojaei, Ph.D.[2], Xinghua Gao, Ph.D.[3], Philip Agee, Ph.D.[4], Abiola Akanmu, Ph.D.[5]

[1] School of Technology, Eastern Illinois University, Charleston, Illinois, United States
[2,3,4,5] Myers-Lawson School of Construction, Virginia Polytechnic Institute and State University, Blacksburg, Virginia, United States

**Abstract**:
Traditional facility management often relies on centralized decision-making structures that limit stakeholder participation, leading to misalignment with occupant needs and decreased satisfaction. This paper proposes a novel blockchain and Decentralized Autonomous Organization (DAO) based framework for community-based facilities management in smart buildings. The framework comprises two key components: a decentralized governance platform that facilitates transparent collective decision-making through blockchain-based voting, and a maintenance management platform with an incentivization mechanism that encourages building occupants to actively contribute to facility upkeep through tokenized rewards. The evaluations of the system included cost analysis, scalability, data security considerations, usability testing, and semi-structured interviews with facility managers and researchers regarding the platform's usefulness, challenges, and adoption potential. The findings demonstrate the framework's potential as a viable incentivization solution for engaging stakeholders in the collective upkeeping and improvement of building infrastructure.

## 1. Introduction

Facilities management (FM) is a multidisciplinary field that encompasses the management of physical assets, services, and resources within the built environment (Nielsen et al. 2016). According to the International Facility Management Association (IFMA), facility management (FM) is an organizational function that combines four key elements—people, place, processes, and technology—with the built environment to enhance individuals' quality of life and increase the efficiency of facilities (IFMA 2014). The traditional FM operations in a built environment are typically operated on centralized organizational structures, where decision-making power typically resides among a few individuals such as building facility managers (Xu et al. 2020). This centralized approach, while initially designed to streamline decision-making, could hinder transparency, and misalign with the building occupant's interests. Previous studies have revealed dissatisfaction among building occupants due to the lack of participation in FM-related decisions that could impact their living environment and experiences (Leaman and Bordass 2001). In response to these challenges, researchers in the built environment domain have emphasized the importance of community-based facility management (CbFM) (Alexander and Brown 2006; Adewunmi et al. 2023), a participatory approach that fosters democratized and socially inclusive facility management practices that prioritize the diverse needs and perspectives of related stakeholders (Tammo and Nelson 2014). CbFM involves the collective participation of community members in managing and maintaining their facilities or infrastructure. This approach not only distributes the decision-making power but also encourages the engagement of all community members in shaping the management and operation of the shared facility (Tammo and Nelson 2012). However, while the CbFM framework decentralizes the decision-making process, its current coordination mechanisms and incentivization system are still rooted in a centralized structure. This centralization can undermine trust and efficiency, as it relies on traditional methods of communication and record-keeping that are not inherently transparent or secure. Without a secure, decentralized system, there can be a lack of trust, transparency, and accountability among stakeholders regarding how decisions are being made and resources allocated.

The advent of blockchain technology (Perera et al. 2020) and decentralized autonomous organization (DAO) (Wang et al. 2019a) presents a potential solution to address the aforementioned challenges in traditional community-based facility management. Blockchain technology offers a decentralized system with a secure ledger that records all transactions and activities in a transparent and immutable manner which could enhance trust and accountability among stakeholders interacting in the CbFM system. Blockchain tokenization also introduces a novel approach to incentivization with the creation of temper-proof digital rewards (e.g. Non-Fungible Tokens and Fungible tokens (Tian et al. 2020)) for the stakeholder's contributions to the CbFM-related activities. Additionally, smart contracts (Khan et al. 2021), a key feature of blockchain, can automate action and enforce agreements within the community without the need for centralized intermediaries. In addition, DAO could enhance the decentralized nature of CbFM by distributing governance among all community members through a decentralized voting system. DAO is a digital and community-driven entity running on a blockchain network that functions transparently and autonomously with democratic and collective decision-making capabilities among its members while having its fundamental operations adhere to rules written in the smart contract code (Singh and Kim 2019). Therefore, the integration of blockchain technology and DAO can potentially enhance the transparency, accountability, and decentralization of stakeholder involvement in CbFM by leveraging the inherent properties of blockchain technology and the DAO-enabled collective and decentralized decision-making capabilities. The DAO-based governance platform will facilitate secure and auditable decision-making processes, while the tokenized incentive mechanism will encourage and reward community contributions to facility maintenance, reporting, and improvement.

This paper proposes an innovative blockchain and Decentralized Autonomous Organization (DAO) based framework for community-based facilities management in smart buildings. The specific objectives of this study are: (1) To examine how decentralized autonomous organizations can enable transparent and collective decision-making in community-based facility management for smart buildings by developing the DAO-based decentralized governance platform. (2) To explore how tokenized incentive systems leveraging blockchain technology can encourage active participation and contributions from stakeholders in CbFM processes. (3) To propose a novel framework that integrates a DAO-based governance platform with an incentivization system for community-based facility management in smart buildings. (4) To implement a full-stack decentralized application (DApp) that facilitates user interactions, voting processes, and incentive distribution within the proposed framework. (5) To conduct a real-world case study in a smart building environment, evaluating the usability, inclusiveness, and decentralization aspects of the developed system through user studies and feedback.

The remainder of this paper is structured as follows: Section 2 provides a comprehensive review of the current practices in facilities management, motivation, and challenges of community-based facilities management, followed by an introduction to the relevant concepts of DAOs and blockchain technology and why there are suitable to address the aforementioned problems. Section 3 outlines the research methodology employed in this study. Section 4 presents the framework of the proposed DAO and blockchain-based CBFM system. Section 5 provides the implementation and prototype of the proposed system. Section 6 describes the evaluation and validation of the system. Then discussion of the findings, implications, limitations of the research, and future research directions is made in section 7. Finally, the conclusion is presented in section 8.

## 2. Departure

In this section, we first explore the motivation behind the concept of community-based facility management and examine its current practices as well as its limitations and challenges. Second, we investigate the potential of decentralized autonomous organization, blockchain, and tokenization through the existing literature. Third, we demonstrate how the DAO's decentralized governance, blockchain inherent security feature, and token-based incentivization framework can address the issue in CbFM.

### 2.1. Toward community-based facility management in built environment

Facilities management (FM) is recognized as the key process by which an organization oversees its buildings, personnel, systems, and support services to ensure alignment with its core business objectives and needs (Chotipanich 2004). FM encompasses a wide range of services and processes essential for the efficient functioning of buildings and infrastructure. It plays a crucial role in ensuring the operational efficiency, safety, and sustainability of buildings and infrastructure (Okoro 2023). However, the traditional FM process has been centralized and managed by a designated team or group of personnel with decision-making authority residing primarily with the facilities manager or management team. Different studies have identified the lack of effective stakeholder participation and engagement as a significant issue in traditional FM practices (Leung et al. 2012). This top-down approach has been associated with several challenges and limitations which can sometimes lead to inefficiencies and user dissatisfaction. Stakeholder engagement is crucial for effective FM, as it allows for the incorporation of diverse perspectives, needs, and experiences from building occupants, tenants, and the surrounding community (Støre-Valen and Buser 2018). Without adequate stakeholder involvement, FM decisions may not fully align with the priorities and preferences of those who interact with the built environment on a daily basis, leading to suboptimal outcomes, decreased occupant satisfaction, and potential conflicts of interest among stakeholders. The traditional FM practices often lack effective communication channels and mechanisms for stakeholders to provide input and feedback (Sedhom et al. 2023). This limitation hinders the ability to gather valuable insights and knowledge from those directly impacted by FM decisions, ultimately leading to inefficiencies, and missed opportunities for improvement.

The concept of community-based facility management (CBFM) has emerged as a more inclusive and participatory approach to FM. CbFM explores opportunities to develop a socially inclusive approach to FM (Hasbullah et al. 2010b). According to Alexander and Brown (2006), CbFM involves managing facilities and services in a way that reflects the community and environment, aiming to empower local communities, spread economic benefits, improve quality of life, and promote local economic development. CbFM recognizes that building occupants, tenants, and community members possess valuable insights and knowledge about the built environment's functionality, efficiency, and overall user experience. By actively involving these stakeholders in decision-making processes, CbFM aims to foster a sense of collective ownership, enhance occupant satisfaction, and promote sustainable practices within the built environment (Tammo and Nelson 2012). The primary motivation behind CbFM is to create a more user-centric and responsive approach to FM, ensuring that the built environment is managed and maintained in a way that meets the diverse needs and preferences of its occupants.

### 2.2. Current Practice and limitation of Community-based Facility Management

To date, various studies have applied the concept of community-based facilities management in practice in the built environment, aiming to enhance occupant satisfaction, resource allocation efficiency, and social inclusivity. Hasbullah et al. (2010), emphasize the social inclusiveness of Community Based Facility Management (CbFM) by involving local school committees in the management and improvement of school facilities. Moghayedi et al. (2024) also explore the potential of implementing CbFM principles to address safety and security concerns on university campuses. In another study on heritage building revitalization, Hou and Wu (2019) demonstrate the effectiveness of Community-based Facilities Management (CbFM) in including diverse stakeholders, such as visitors, tenants, operational staff, and public and private sector entities in the decision-making process. This inclusive approach ensures that revitalized buildings are functional, creatively designed, and meet the needs of all parties. In addition, the study by Mugumya (2013) highlights the significance of decentralized and socially inclusive approaches to achieve efficient resource allocation, particularly in the context of water resource management. In another study, Abowen-Dake and Nelson (2013) also proposed the use of CbFM approach in the management and improvement of the Library's facilities. The study highlights the potential benefits of active community participation in various aspects, such as assessing needs, drafting specifications in Service Level Agreements (SLAs), and suggesting improvements to services.

However, despite the advantages provided by community-based facility management, there are still a few challenges and limitations that hinder its full potential for effective implementation. Although the CbFM framework aims to distribute decision-making power, its existing coordination mechanisms remain reliant on centralized, Web 2.0 technologies. This centralization poses challenges to trust and efficiency, as it depends on conventional communication and record-keeping methods that lack inherent transparency and security. For instance, research conducted by Sedhom et al. (2023) sheds light on two primary challenges faced in community-based facility management: information management and stakeholder engagement. The main challenge in information management is the unreliable data source. Traditional centralized systems often struggle to maintain the integrity and accuracy of data with the lack of transparency. Furthermore, they mentioned that the main challenges in stakeholder engagement are a lack of trust and transparency in communication between the involved parties. In addition, research by Abowen-Dake and Nelson (2013) has found that one of the main barriers to implementing effective community-based facilities management is the lack of people's willingness to participate in the decision-making process. Therefore, it's also important to seek solutions that encourage greater involvement from community members. One promising approach is incentivization, wherein individuals are motivated to participate through various rewards or recognition mechanisms. These challenges and limitations highlight the need for innovative solutions that can enhance data integrity, incentivization framework, as well as the transparency and efficiency of the coordination process within CbFM.

### 2.3. Blockchain and decentralized autonomous organization

#### 2.3.1. Blockchain technologies

Blockchain is a digital public ledger that has all its data documented and stored in a transparent, and tamper-resistant manner in the decentralized network. The blockchain is built over a peer-to-peer network that distributes the workload among all peers (Lee et al. 2021). This decentralized nature is a core feature of blockchain that distinguishes it from traditional centralized systems, where data and control are concentrated in a small group of entities. Instead, blockchain leverages a distributed network of nodes to collectively validate and record transactions through a consensus mechanism (e.g. Proof of Work, Proof of stake) that ensures all nodes agree on the validity of data before it is added to the immutable chain (Perera et al. 2020). In a blockchain database, information is organized into blocks which are interconnected to form a chain. The newly created block after data validation is appended to the blockchain network in a chronological and immutable fashion using hash codes and forms a longer chain (Yaga et al. 2018). This architecture makes it challenging for anyone to modify the content of a block since any alteration made to a block will render all of the succeeding blocks invalid (El Ioini and Pahl 2018). This structure also provides data traceability by cryptographically linking each new block of data to the previous one, forming an auditable and tamper-resistant trail of records (Kiu et al. 2022).

Modern blockchain networks such as Ethereum extended their applications beyond cryptocurrency transactions and data security with the introduction of smart contracts (Wang et al. 2021). Smart contracts are self-executing computer programs that offer self-enforcing and secure task execution capabilities based on a decentralized consensus (Khan et al. 2021). This capability has paved the way for the creation of decentralized applications or DApp (Cai et al. 2018), software applications that operate on a decentralized network with blockchain technology.

#### 2.3.2. Blockchain-based incentivization in construction

The creation of smart contracts also facilitates incentivization and tokenization processes, enabling the creation of digital assets and reward mechanisms within the blockchain ecosystems (John et al. 2023). Tokenization involves creating digital tokens that represent ownership or participation in real-world assets or services. There are two primary types of blockchain-based tokens: fungible tokens and non-fungible tokens (NFTs) (Tian et al. 2020). Fungible tokens, such as cryptocurrencies like Bitcoin and Ethereum, are identical and can be exchanged on a one-to-one basis. NFTs, on the other hand, are unique digital assets that represent ownership of a specific item or piece of content, making them ideal for representing reputational badges or collectibles (Naderi et al. 2024). In blockchain-based incentivization, users can earn tokens as rewards for their contributions, such as through badges or reputation points. These tokens can be fungible, providing a tangible financial incentive, or non-fungible, serving as unique markers of achievement, reputation, and status within a community (Voshmgir and Zargham 2019). Researchers have also explored the use of blockchain-based incentivization in the construction domain. For instance, Naderi et al. (2023) introduced a blockchain-enabled incentivization mechanism for construction safety, where smart contracts automatically distribute fungible tokens (FTs) and non-fungible tokens (NFTs) based on safety compliance. The system leverages computer vision to analyze visual data from construction sites, generating safety performance reports that are evaluated using a Decentralized Oracle Network (DON). Additionally, Hunhevicz et al. (2022) investigated the integration of digital building twins with blockchain-based smart contracts for performance-based contracting. Their research introduced a technical architecture that connects digital twins, IoT sensors, and blockchain to automate performance evaluation and reward stakeholders based on real-time performance data. Another study by Hunhevicz et al. (2020) explored blockchain-based incentivization for high-quality data management in construction projects. Their research proposed a smart contract system on an Ethereum-based blockchain to encourage the creation and maintenance of high-quality data sets throughout a construction project's lifecycle.

#### 2.3.3. Decentralized autonomous organization

A decentralized autonomous organization is a digital and community-driven entity running on a blockchain network that functions transparently and autonomously with democratic and collective decision-making capabilities among its members while having its fundamental operations adhere to rules written in the smart contract code (Rikken et al. 2023). The decentralized autonomous organization (DAO) represents a novel organizational paradigm that fundamentally departs from conventional centralized structures. DAOs are underpinned by three core pillars: decentralization, autonomy, and automation (Wang et al. 2019a). Firstly, in contrast to hierarchical top-down management, DAOs operate through a decentralized peer-to-peer network of nodes on the underlying blockchain, eliminating the centralized governing authority (Singh and Kim 2019). Secondly, DAOs are designed as self-governing autonomous entities where governance occurs through the collective participation and voting input of community members incentivized by a token-based mechanism (Santana and Albareda 2022). DAO's proposals are initiated and approved through the decentralized democratic process. Finally, building upon blockchain's immutability and transparency, smart contracts encoded with predefined rules and regulations enable the automation of the DAO's organizational operations and transactions (Dwivedi et al. 2021). The interconnection between the technological characteristics of blockchain and Decentralized Autonomous Organizations (DAOs) is depicted in Fig. 1. Blockchain technology establishes crucial technological foundations by offering core security features such as secure and immutable record-keeping and smart contracts. Simultaneously, DAOs can provide an additional overlaying organizational layer with decentralized coordination, voting mechanisms, and incentive models. By synergistically combining blockchain's fundamental features with its inherent governance mechanism, DAOs present novel opportunities for creating a governing entity and decentralized organizational structures that can potentially enhance the transparency and efficiency of the coordination process in the CbFM through the blockchain-based decentralized incentivization framework and decentralized governance (Ly and Shojaei 2025).

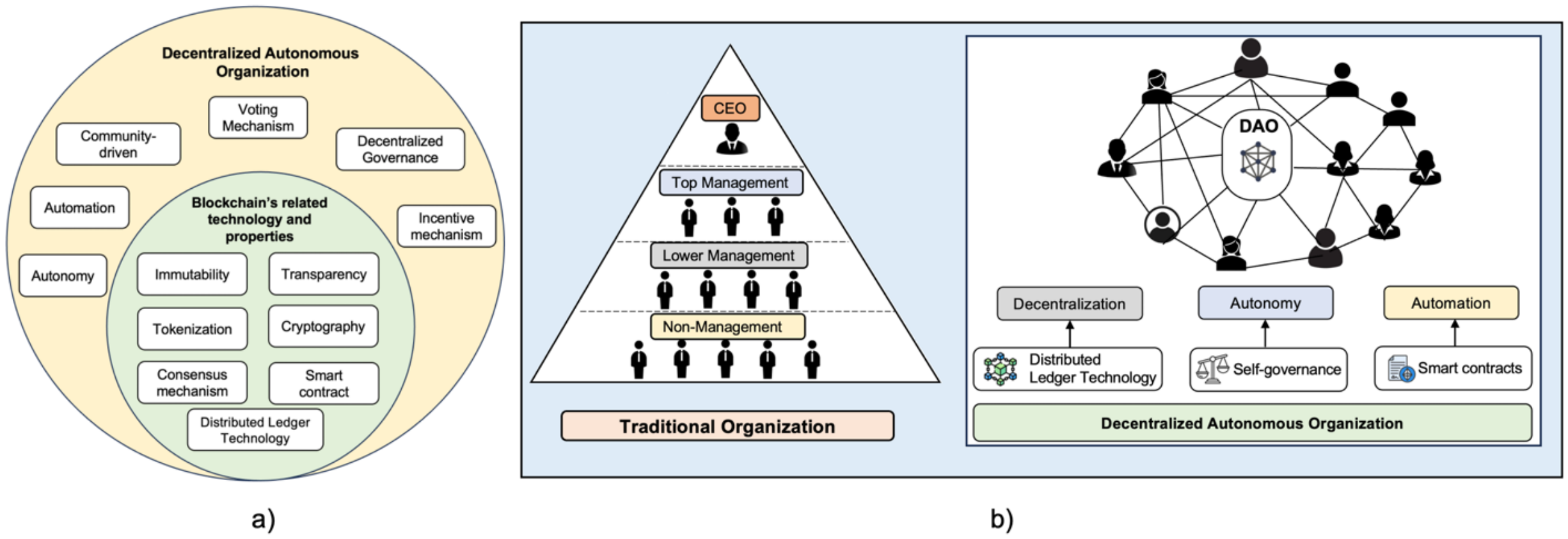

**Fig. 1** Decentralized autonomous organization. a) DAO's Technical properties b) The difference in structures between the traditional Organization and DAO

#### 2.3.4. DAO in the AEC industry

Over the past few years, multiple studies have demonstrated DAO's capabilities in facilitating decentralized coordination of project management processes. In their study, Spychiger et al (2023), developed a Decentralized Autonomous Project Organization (DAPO), a DAO-based project management platform based on the Ethereum network. Using the platform-based DAO approach (Aragon), Darabseh and Poças Martins (2023) have demonstrated a DAO use case in real-world construction practice by creating a prototype of a decentralized governance system for construction projects. In another work by Dounas et al. (2022), the integration of the stigmergic principle, blockchain immutability, and DAO's decentralized governance was proposed to foster collaboration and collective ownership in architectural design. The proposed system, ArchiDAO (2022), essentially operates as a decentralized design studio based on blockchain where any designer can join and work collaboratively on the project. In addition, Ly et al. (2024) also proposed a conceptual framework that integrates digital twin and DAO framework for smart building facilities management. Another study by Ly et al. (2024) further develops this concept by designing and prototyping a Decentralized Autonomous Building Cyber-Physical System framework that incorporates DAOs, Large Language Models (LLMs), and digital twins to create a self-managed, operational, and financially autonomous building system. Their research validates the framework through a full-stack decentralized application and an LLM-based AI assistant, demonstrating its feasibility in real-world building management scenarios, such as AI-assisted facility control and DAO-based revenue and expense management.

### 2.4. Research gaps and scope of the study

The research on DAO and its application in the previous section demonstrates the feasibility and effectiveness of DAOs in enabling decentralized coordination within different research domains including project and construction management processes. However, there is a notable knowledge gap in the understanding of decentralized governance in the context of physical infrastructure such as smart building facility management. While the work by Ly et al. (2024) provided a conceptual framework for DAO application in facilities management, there is a lack of empirical studies that have fully explored and implemented this concept. Furthermore, the research on DAO governance applications specifically for community-driven facility management in smart buildings remains largely unexplored.

This study aims to bridge this gap by developing a comprehensive framework for decentralized community-based facility management (CbFM) with an integrated incentivization mechanism. The scope of the research will focus on Operations and Maintenance (O&M) which is one of the eleven core aspects of facility management defined by the International Facility Management Association (IFMA) (IFMA 2022). In addition to the theoretical and technical components, the study will conduct simulated case study and user studies to evaluate the practical application and effectiveness of the developed framework and DApp.

## 3. Research Methodology

This study adopts the Design Science Research (DSR) methodology, a problem-solving paradigm aimed at creating innovative artifacts (e.g. algorithms, prototypes, frameworks, or models) to solve real-world problems and contribute to the body of knowledge (Peffers et al. 2012). The DSR approach has been widely used by researchers in the construction industry for developing blockchain-related applications, including blockchain frameworks for construction cost management (Cheng et al. 2023), lightweight blockchain-as-a-service frameworks to enhance BIM security (Tao et al. 2023), and decentralized material management systems for construction projects (Basheer et al. 2024). Fig. 2 presents the research stage with the corresponding DSR process to develop a DAO-based decentralized governance platform and blockchain-based incentivization system for community facilities management. The DSR process involves six iterative steps:

(1) Identification of problem and motivation. A literature review was conducted in section 2 to explore the challenges and limitations of traditional FM practices, particularly in the context of community-based facility management (CbFM). The result led to the initial motivation of this study, a lack of decentralized and incentivization frameworks that enable transparent decision-making, effective coordination, and active stakeholder engagement in the CbFM processes.

(2) Definition of Objectives. The primary objective of this study is to develop a decentralized governance platform and incentive mechanism for community-based facility management in smart buildings. Decentralized autonomous organization and blockchain technology are identified as the main components to achieve this objective.

(3) Design and Development. The design and development phase involves the identification and integration of various components and modules to create the proposed system. Key components include (i) Decentralized Governance: A decentralized governance model based on a DAO, enabling collective decision-making, and voting processes among stakeholders. (ii) Tokenization: Utilization of blockchain-based tokenization mechanisms, including fungible tokens for incentivization and non-fungible tokens (NFTs) for recognizing and tracking contributions. (iii) Decentralized Web Applications: Development of decentralized applications (DApp) to facilitate user interactions with the platform.

(4) Demonstration. The developed CBFM system will be deployed on an Ethereum test network to simulate various scenarios and validate its functionality. This includes testing the DAO governance processes, tokenized incentive mechanisms, and user interactions through the Dapp.

(5) Evaluation. Both quantitative measures and qualitative assessments of the framework will be provided.

(6) Communication. A prototype demonstration will be conducted to receive feedback from relevant stakeholders, including facility managers, building owners, and occupants. The design and development process of the system and evaluation results will be published in academic journals.

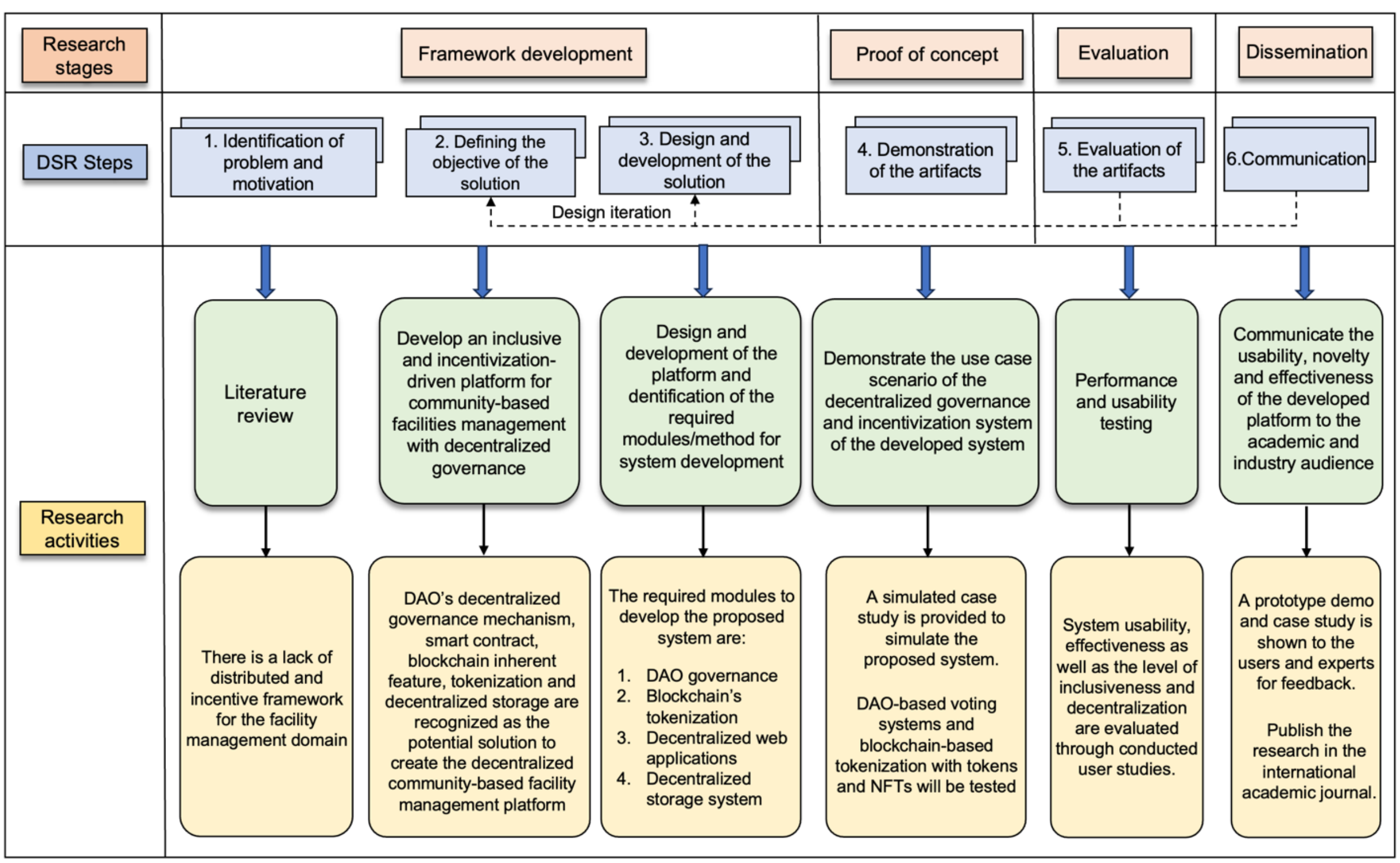

**Fig. 2** Design Science Research-driven research flow

## 4. Proposed decentralized community-based facility management framework

### 4.1. Framework overview

The primary objective of this framework is to encourage occupants and related stakeholders to actively contribute to the upkeep, improvement, and sustainability of the shared building infrastructure through the blockchain-based incentivization scheme by distributing the fungible tokens and non-fungible tokens (reputational tokens) through the proposed decentralized application. Fig. 3 provides a high-level overview of the proposed framework and its comprised components which will be further discussed in the following sections. The framework comprises three primary components: a physical component, represented by the building infrastructure, and two cyber components, namely the Decentralized Governance Platform and the Maintenance Management Platform, which are the main modules in the proposed decentralized applications.

The Maintenance Management Platform facilitates occupants in submitting maintenance requests, work orders, feedback, and sustainability initiatives through the DApp. Occupants can provide textual descriptions, locations, and multimedia attachments from the building infrastructure within their proposal submissions. The incentivization component in the governance platform introduces an incentive mechanism through blockchain-based tokens. Occupants who submit valid and relevant reports or participate in the voting process can earn fungible reward tokens and non-fungible reputation/reward tokens (NFTs) based on the ERC-721 token standard. The incentivization logic is encoded in smart contracts to ensure fair and transparent distribution of tokens based on the established rules and conditions.

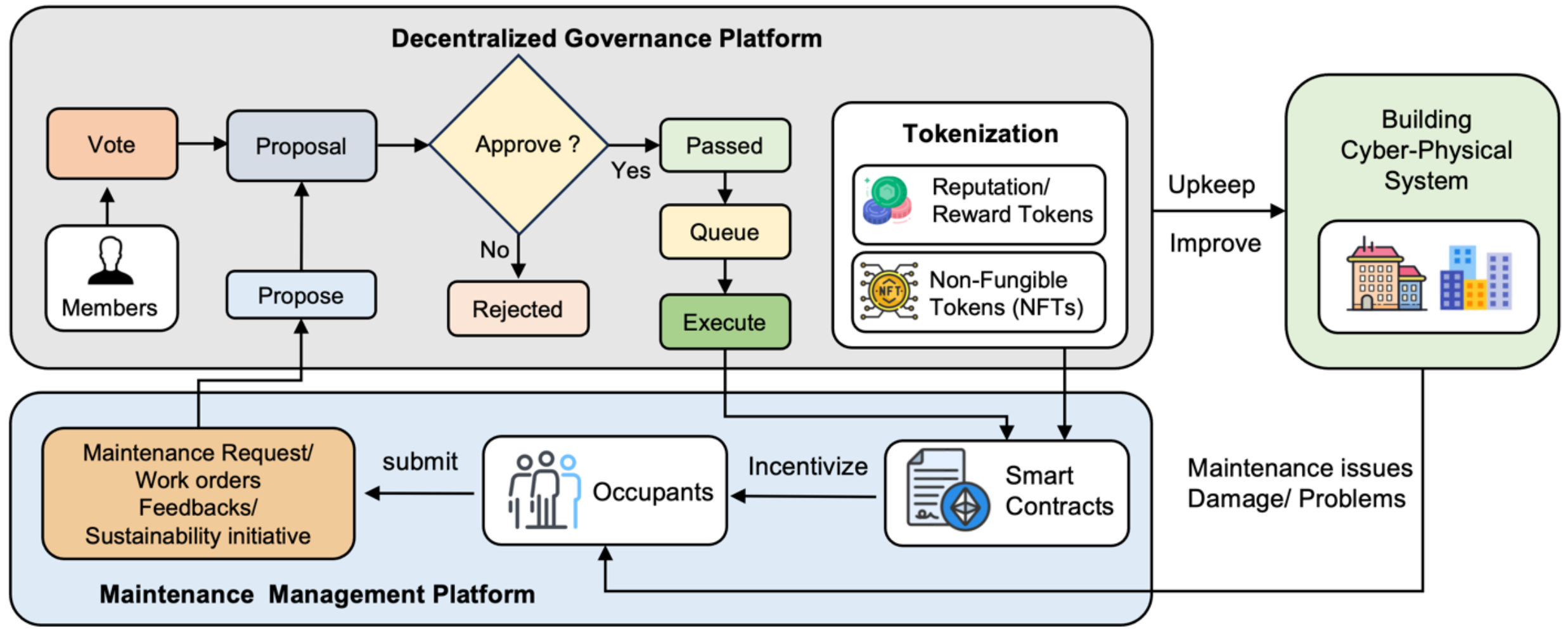

**Fig.3** Overview of the decentralized governance and incentivization framework for community-based facility management

### 4.2. Decentralized governance platform

The decentralized governance platform serves as the core decision-making and coordination hub for the proposed community-based facility management system. The architecture of the platform's framework including its governance process and key functionality are illustrated in Fig. 4. One of the key responsibilities of the Decentralized Governance Platform is to approve and oversee the incentivization processes. Members can collectively approve the incentivization mechanisms, ensuring fairness and transparency in rewarding occupants for their contributions. The Decentralized Governance Platform also plays a crucial role in treasury management and the management of token supply. It oversees the minting of fungible tokens and non-fungible tokens (NFTs) used for incentivization and reputation purposes. Members can collectively decide on budget allocations, enabling the platform to fund new initiatives, policies, or projects related to building maintenance and improvement.

Voting mechanisms are central to the functioning of the decentralized governance platform. Once DAO is first deployed on the blockchain network, a specified amount of governance tokens will be minted and distributed to key members corresponding to their roles and responsibilities. They will be granted a higher number of governance tokens, reflecting their expertise and decision-making authority within the platform. These tokens will provide them with more significant voting power and governance rights compared to regular occupant members. The occupants can gradually accumulate governance tokens and increase their voting influence by actively participating in the system, submitting valid maintenance requests, providing valuable feedback, and contributing to the upkeep and improvement of the building infrastructure. This approach ensures that the platform maintains a balance between the expertise and responsibilities of key stakeholders while fostering a sense of ownership and incentivizing active participation from occupants.

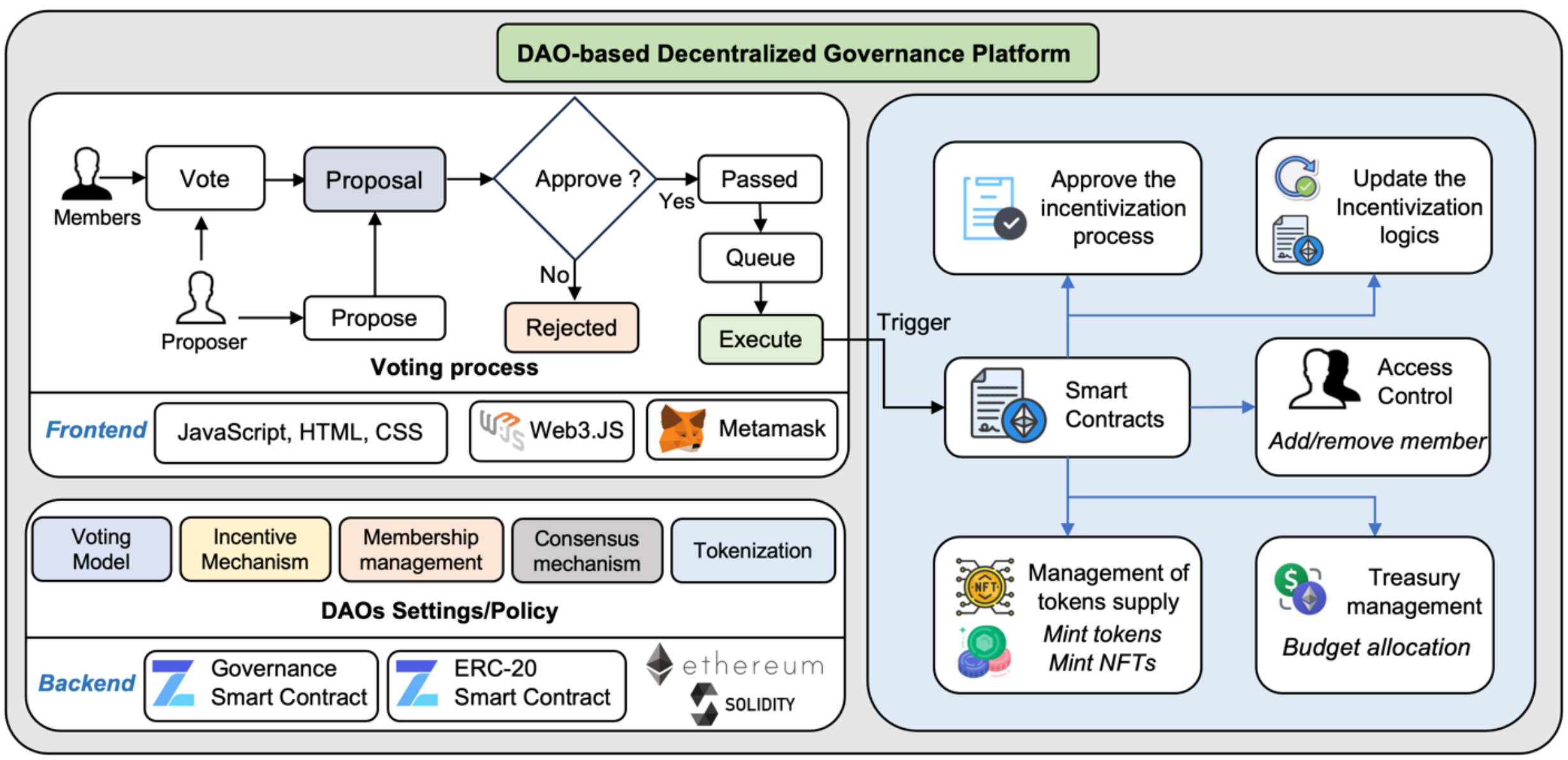

**Fig.4** Framework of the decentralized governance platform

### 4.3. Maintenance Management Platform

The Maintenance Management Platform facilitates the submission of maintenance requests, work orders, feedback, and sustainability initiatives by building occupants. Its primary objective is to provide a user-friendly interface for occupants to report issues and improvement feedback to the main decentralized governance platform to upkeep and improve their building infrastructure**.** The framework architecture of the maintenance submission system as well as its relationship with the occupants and the decentralized governance platform are illustrated in Fig. 5.

The workflow begins with maintenance issue-related data collection where occupants capture images or videos of problems from their building infrastructure using their mobile devices or cameras. Occupants can then access the maintenance platform's user interface and upload the captured multimedia attachments along with descriptive text explaining the issue or providing feedback. This seamless process aims to enhance the user experience thereby encouraging the platform usage and active occupant participation in the maintenance and enhancement of the building environment. Once the related document is submitted to the platform, the multimedia data (e.g. image and video) are automatically uploaded to a decentralized storage system, such as the Interplanetary File System (IPFS). The IPFS generates a unique Content Identifier (CID) for each uploaded file, thereby enabling transparent and immutable access to the submitted content. By leveraging decentralized storage, the framework also ensures the platform's scalability by avoiding the high costs associated with storing large files directly on the blockchain. The Maintenance Management Platform then combines the occupant's descriptive text with the CID of the uploaded multimedia data to create a finalized DAO proposal which will be used for submission to the Decentralized Governance Platform. Once the proposal is submitted, DAO members on the Decentralized Governance Platform can review the proposal details, using the occupant's textual description and the multimedia content retrieved by the associated CID, enabling them to make informed decisions during the voting process.

### 4.4. Incentive mechanism

The incentive mechanism is one of the core components within the proposed decentralized framework for community-based facility management. This framework leverages blockchain-based tokenization, utilizing both fungible tokens and reputational non-fungible tokens to incentivize and reward occupant actions and contributions within the community-based facility management system. The dual-token system used in the platform ensures that participants are rewarded not just for their actions but also for their commitment and reputation, which could effectively motivate more user engagement and contribution to the building maintenance and governance processes over time. Occupants who submit valid and relevant maintenance issues or tasks will be rewarded with CBFMT (Community-based facility management tokens) once their proposal has been approved by the decentralized governance platform. In addition to the fungible token rewards (CBFMT), the platform introduces a reputation system based on non-fungible tokens, namely CBFMNFT (Community-based facility management non-fungible tokens). These reputation tokens, or CBFMNFTs, serve as a metric of the occupants' long-term contributions and standing within the community. These rewarded tokens can potentially be utilized in various ways in the real world depending on the specific nature and requirements of the building or community.

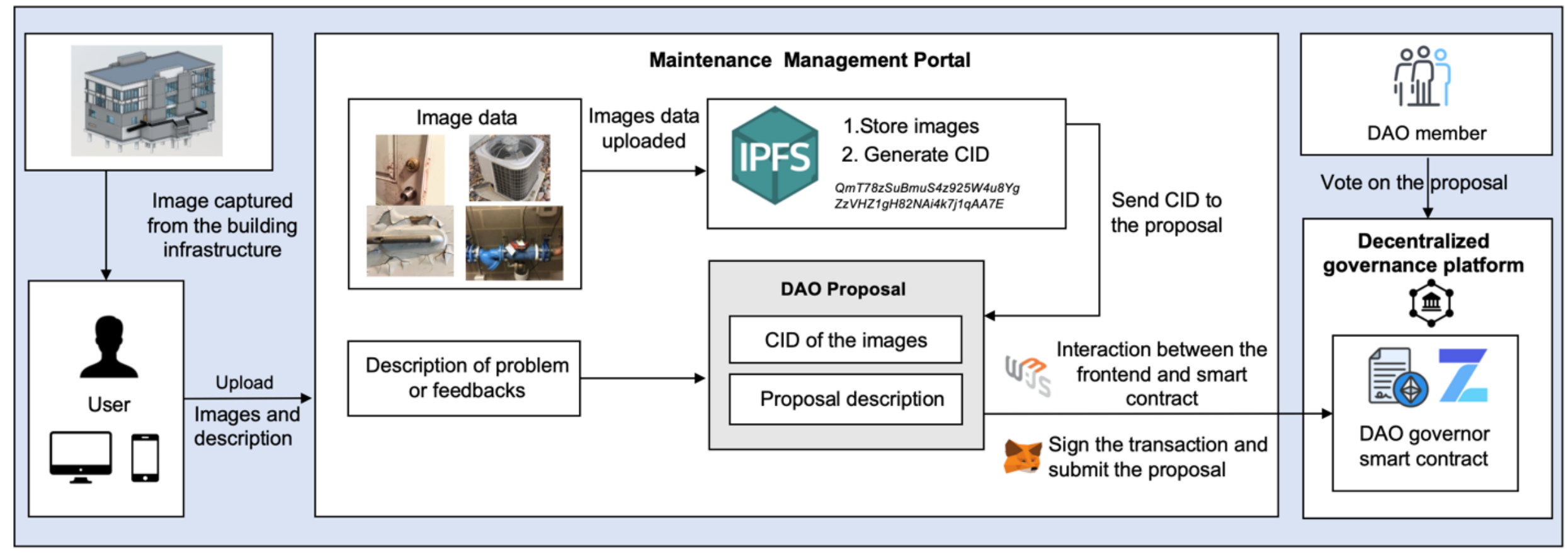

**Fig.5** Framework architecture of the maintenance submission system

## 5. Proof of concept

In this section, a case study with the developed prototypes is used to validate the viability and functionality of the framework. The tools, coding languages, and development environments employed for each module of the prototypes are summarized in Table 1.

**Table 1** Tools used for prototype development

| Tasks | Programming language (packages) | Development environment |
|---|---|---|
| Frontend web pages development | React JS | Visual Studio Code |
| Smart contract development | Solidity | Brownie |
| Digital building twin | JavaScript (Autodesk API) | Visual Studio Code |
| IoT sensors and smart home device | Python | Visual Studio Code |
| Interaction between Dapp and smart contract | JavaScript (web3.js API) | Visual Studio Code |

### 5.1. Development of Dapp backend

#### 5.1.1. Smart contract design and development

The decentralized governance DApp comprises five main smart contracts including the DAO governor contract, time lock contract, Governance tokens contract, CbFM NFT contract, and the incentive logics contract. We utilized the base smart contracts from the OpenZepplin library including the DAO governor contracts, as well as the ERC-20 and ERC-721 tokens contracts. The design of the five smart contracts and the relationship between their function of the roles of actors in the proposed framework are illustrated in Fig. 6.

The DAO Governor contract is designed to facilitate the core governance processes within the platform. It manages proposal submissions, voting mechanisms, and execution of approved decisions (Fig. 6). It also handles the vote-counting process and determines whether a proposal has passed based on predefined quorum and majority requirements. The DAO Governor contract in this study inherits from several OpenZepplin base contracts. For instance, The Governor base contract provides the core functionality for proposal creation and execution. GovernorTimelockControl contract adds a security layer by delaying proposal execution. The GovernorVotes contract ensures that voting power is derived from the ERC-20 governance tokens. GovernorCountingSimple implements a straightforward vote-counting mechanism, while GovernorVotesQuorumFraction enforces a quorum based on a fraction of the total token supply. In addition, GovernorSettings allows the configuration of governance parameters like voting delay and voting period.

The governance tokens contract is designed to manage the distribution and management of the platform's fungible tokens (CBFMT) (Fig. 8.a). Which represents voting power within the DAO-based governance platform. This contract handles the minting of new governance tokens upon deployment and as needed. It serves as the medium for both the governance and transactional activities within the system. The Governance Tokens contract inherits from the ERC20Votes base contract. ERC20Votes contract extends the standard ERC20 token with voting and delegation capabilities. This allows token holders to either vote directly or delegate their voting power to other addresses. This proposed platform facilitates the delegation of voting power by allowing the allocation of tokens to key members and participants based on their roles and contributions, and the transfer of tokens between members. In addition, the CbFM_NFT contract is created to manage the platform's non-fungible tokens (CBFMNFTs), which represent the reputation and long-term contributions within the community (Fig. 9.b). It inherits from several base contracts, including ERC721, ERC721Burnable, ERC721URIStorage, and IERC721Enumerable which provide the core functionality for creating, managing, removing, and tracking the NFTs. These functionalities allow DAO members to incentivize the participants for their actions and contributions to the community by minting them the NFTs and facilitating the trading or exchange of NFTs among participants.

Furthermore, the Timelock contract serves as a security measure in enforcing the governance decisions by introducing a mandatory delay between the approval of a proposal and its execution (Fig. 8.b). It queues approved proposals for a specified delay period and executes them only after this period has passed. This ensures that all stakeholders have adequate time to review the approved decisions and react if necessary, preventing rash or malicious actions. Moreover, the Incentive logic contract was developed to manage the reward distribution system (Fig. 9.a). DAO members can propose changes to several tokens and NFT for the voting participation, the participant's successful proposal as well as the exchange rate between the fungible (CBFMT) and non-fungible (CBFMNFT). This contract interacts with both the governance tokens contract, CbFM_NFT as well as the Governor contract to manage the tokens distribution process.

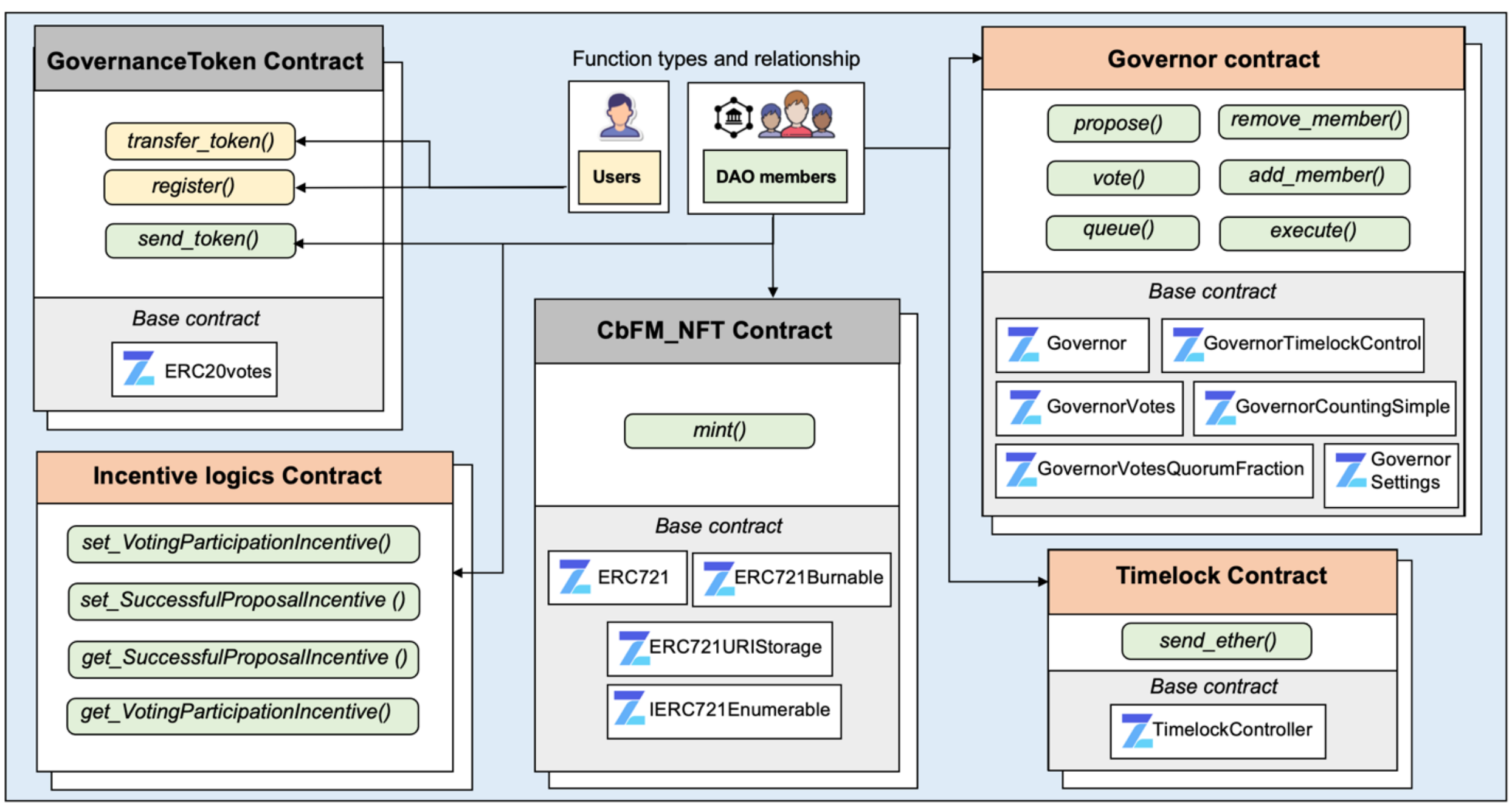

**Fig.6** Design of the smart contracts and their relationship between actors in the systems

```solidity
contract DABCPSGovernor is
    Governor,
    GovernorSettings,
    GovernorCountingSimple,
    GovernorVotes,
    GovernorVotesQuorumFraction,
    GovernorTimelockControl
{
    // Proposal Counts
    uint256 public s_proposalCount;
    address[] public members;
    mapping(address => bool) public isMember;
    constructor(
        IVotes _token,
        TimelockController _timelock,
        uint256 _votingDelay,
        uint256 _votingPeriod,
        uint256 _quorumPercentage
    )
        Governor("Governor")
        GovernorSettings(
            _votingDelay, /* 1 => 1 block */
            _votingPeriod, /* 300 blocks => 1 hour */
            0 /* 0 => Because we want anyone to be able to create a proposal */
        )
        GovernorVotes(_token)
        GovernorVotesQuorumFraction(_quorumPercentage) /* 4 => 4% */
        GovernorTimelockControl(_timelock)
    {
        s_proposalCount = 0;
        members.push(msg.sender);
        isMember[msg.sender] = true;
    }
    fallback() external payable {}
    modifier onlyMember() {
        require(isMember[msg.sender], "Only members can call this function");
        _;
    }
    function sendEther(address payable receiver, uint256 amount) external {
        require(address(this).balance >= amount, "Insufficient balance in the contract");
        receiver.transfer(amount);
    }
    function getMembers() external view returns (address[] memory) {
        return members;
    }
    // The following functions are overrides required by Solidity.
    function addMember(address newMember) external onlyMember {
        require(!isMember[newMember], "Address is already a member");
        members.push(newMember);
        isMember[newMember] = true;
    }
    function removeMember(address member) external onlyMember {
        require(isMember[member], "Address is not a member");

        // Find the index of the member in the array
        uint256 index;
        for (uint256 i = 0; i < members.length; i++) {
            if (members[i] == member) {
                index = i;
                break;
            }
        }
        // Swap with the last element and then remove the last element to maintain order
        members[index] = members[members.length - 1];
        members.pop();
        isMember[member] = false;
    }
    function getMemberLength() external view returns (uint256) {
        return members.length;
    }
    function votingDelay()
        public
        view
        override(IGovernor, GovernorSettings)
        returns (uint256)
    {
        return super.votingDelay();
    }
```

```solidity
    function votingPeriod()
        public
        view
        override(IGovernor, GovernorSettings)
        returns (uint256)
    {
        return super.votingPeriod();
    }
    function quorum(uint256 blockNumber)
        public
        view
        override(IGovernor, GovernorVotesQuorumFraction)
        returns (uint256)
    {
        return super.quorum(blockNumber);
    }
    function state(uint256 proposalId)
        public
        view
        override(Governor, GovernorTimelockControl)
        returns (ProposalState)
    {
        return super.state(proposalId);
    }
    function propose(
        address[] memory targets,
        uint256[] memory values,
        bytes[] memory calldatas,
        string memory description
    ) public override(Governor, IGovernor) returns (uint256) {
        s_proposalCount++;
        return super.propose(targets, values, calldatas, description);
    }
    function proposalThreshold()
        public
        view
        override(Governor, GovernorSettings)
        returns (uint256)
    {
        return super.proposalThreshold();
    }
    function _execute(
        uint256 proposalId,
        address[] memory targets,
        uint256[] memory values,
        bytes[] memory calldatas,
        bytes32 descriptionHash
    ) internal override(Governor, GovernorTimelockControl) {
        super._execute(proposalId, targets, values, calldatas, descriptionHash);
    }
    function _cancel(
        address[] memory targets,
        uint256[] memory values,
        bytes[] memory calldatas,
        bytes32 descriptionHash
    ) internal override(Governor, GovernorTimelockControl) returns (uint256) {
        return super._cancel(targets, values, calldatas, descriptionHash);
    }
    function _executor()
        internal
        view
        override(Governor, GovernorTimelockControl)
        returns (address)
    {
        return super._executor();
    }
    function supportsInterface(bytes4 interfaceId)
        public
        view
        override(Governor, GovernorTimelockControl)
        returns (bool)
    {
        return super.supportsInterface(interfaceId);
    }
    function getNumberOfProposals() public view returns (uint256) {
        return s_proposalCount;
    }
}
```

**Fig. 7** DAO's Governor contract

```
5   pragma solidity ^0.8.7;
6
7   contract GovernanceToken is ERC20Votes {
8       // events for the governance token
9       event TokenTransfered(
10          address indexed from,
11          address indexed to,
12          uint256 amount
13      );
14      // Events
15      event TokenMinted(address indexed to, uint256 amount);
16      event TokenBurned(address indexed from, uint256 amount);
17      // max tokens per user
18      uint256 constant TOKENS_PER_USER = 2000;
19      uint256 constant TOTAL_SUPPLY = 1000000 * 10**18;
20      uint256 public data;
21      // Mappings
22      mapping(address => bool) public s_claimedTokens;
23      // Number of holders
24      address[] public s_holders;
25      constructor(uint256 _keepPercentage)
26          ERC20("BFHTOKEN", "BFHT")
27          ERC20Permit("BFHToken")
28      {
29          uint256 keepAmount = (TOTAL_SUPPLY * _keepPercentage) / 100;
30          _mint(msg.sender, TOTAL_SUPPLY);
31          _transfer(msg.sender, address(this), TOTAL_SUPPLY - keepAmount);
32          s_holders.push(msg.sender);
33      }
34      function sendTokens(address payable receiver, uint256 amount) external  {
35          _transfer(address(this), receiver, amount * 10**18);
36      }
37      function reward(uint256 amount) external  {
38          _transfer(address(this), msg.sender, amount * 10**18);
39      }
40      function getHolderLength() external view returns (uint256) {
41          return s_holders.length;
42      }
43      // Overrides required for Solidiy
44      function _afterTokenTransfer(
45          address from,
46          address to,
47          uint256 amount
48      ) internal override(ERC20Votes) {
49          super._afterTokenTransfer(from, to, amount);
50          emit TokenTransfered(from, to, amount);
51      }
52      function _mint(address to, uint256 amount) internal override(ERC20Votes) {
53          super._mint(to, amount);
54          emit TokenMinted(to, amount);
55      }
56      function _burn(address account, uint256 amount)
57          internal
58          override(ERC20Votes)
59      {
60          super._burn(account, amount);
61          emit TokenBurned(account, amount);
62      }
63  }
```

a)

```
1   //SPDX-License-Identifier: MIT
2
3   pragma solidity ^0.8.9;
4
5   import "@openzeppelin/contracts/governance/TimelockController.sol";
6
7   contract TimeLock is TimelockController {
8       constructor(
9           uint256 minDelay,
10          address[] memory proposers,
11          address[] memory executors,
12          address admin
13      ) TimelockController(minDelay, proposers, executors, admin) {}
14
15      function sendEther(address payable receiver, uint256 amount) external {
16          require(address(this).balance >= amount, "Insufficient balance in the contract");
17          receiver.transfer(amount);
18      }
19
20  }
```

b)

BFHTOKEN Top 100 Token Holders

Source: Etherscan.io

0x1234567890123456789012345678901234567890
0xca36eb5091e0fe1c40005e30d5c4e047b668f3fa
0x13a502ec51b27dfadb174c6d57e06150df762905
0xc9602ab8682443e3e0391aa9b86e0360479b85c2
0xd760a17210b61900f696ad0c21a6e11dc8884198
0xd0bcd44a1f11e96c06abf08f973a775e1c09fece

(A total of 1,000,000.00 tokens held by the top 100 accounts from the total supply of 1,000,000.00 token)

| Rank | Address | Quantity (Token) | Percentage |
|---|---|---|---|
| 1 | 0xd0bcD44A...e1c09FecE | 642,820 | 64.2820% |
| 2 | 0xD760A172...dc8884198 | 317,331.099999999999998821 | 31.7331% |
| 3 | 0xC9602ab8...0479B85C2 | 20,228.900000000000000109 | 2.0229% |
| 4 | 0x13A502ec...0df762905 | 12,000 | 1.2000% |
| 5 | 0xcA36EB50...7b668F3fa | 7,430.000000000000000089 | 0.7430% |
| 6 | 0x12345678...234567890 | 180 | 0.0180% |
| 7 | 0x70997970...d17dc79C8 | 10 | 0.0010% |

c)

**Fig. 8** Governance tokens related smart contracts (a) Governance token contract (b) Timelock controller contract (c) Distribution of tokens on Ethereum Sepolia Testnet

```solidity
// SPDX-License-Identifier: MIT
pragma solidity ^0.8.0;

contract DAOIncentives {
    // Variables to store incentive amounts
    uint256 public votingParticipationIncentive;
    uint256 public successfulProposalIncentive;

    // Address of the DAO admin (for setting initial values, can be transferred if needed)
    address public admin = 0x3aF5647E366fb51C89e4c43Bc8C173dAa018AFf6;

    // Modifier to restrict access to admin functions
    modifier onlyAdmin() {
        require(msg.sender == admin, "Only admin can call this function");
        _;
    }
    // Constructor to initialize the admin and default values
    constructor(uint256 _votingIncentive, uint256 _proposalIncentive) {
        votingParticipationIncentive = _votingIncentive;
        successfulProposalIncentive = _proposalIncentive;
    }
    // Function to set the voting participation incentive
    function setVotingParticipationIncentive(uint256 _newIncentive) external onlyAdmin {
        votingParticipationIncentive = _newIncentive;
    }
    // Function to get the voting participation incentive
    function getVotingParticipationIncentive() external view returns (uint256) {
        return votingParticipationIncentive;
    }
    // Function to set the successful proposal incentive
    function setSuccessfulProposalIncentive(uint256 _newIncentive) external onlyAdmin {
        successfulProposalIncentive = _newIncentive;
    }

    // Function to get the successful proposal incentive
    function getSuccessfulProposalIncentive() external view returns (uint256) {
        return successfulProposalIncentive;
    }
    // Function to transfer admin rights to a new address (if needed)
    function transferAdmin(address _newAdmin) external onlyAdmin {
        require(_newAdmin != address(0), "Invalid admin address");
        admin = _newAdmin;
    }
}
```

a)

```solidity
// SPDX-License-Identifier: MIT
pragma solidity ^0.8.20;

import "@openzeppelin/contracts/token/ERC721/ERC721.sol";
import "@openzeppelin/contracts/token/ERC721/extensions/ERC721URIStorage.sol";
import "@openzeppelin/contracts/token/ERC721/extensions/ERC721Burnable.sol";
import "@openzeppelin/contracts/token/ERC721/extensions/IERC721Enumerable.sol";
import "@openzeppelin/contracts/access/Ownable.sol";

contract MyTokennew is ERC721, ERC721URIStorage, ERC721Burnable, IERC721Enumerable {
    address[] public members;
    mapping(address => bool) public isMember;
    mapping(address => uint256[]) private _ownedTokens;
    mapping(uint256 => uint256) private _ownedTokensIndex;
    mapping(uint256 => uint256) private _allTokensIndex;
    uint256[] private _allTokens;
    uint256 private _totalSupply; // Track the total supply of tokens
    constructor () ERC721("MLSOCNFT", "MLSOCNFT") {
        members.push(msg.sender);
        isMember[msg.sender] = true;
        _totalSupply = 0; // Initialize the total supply
    }
    function safeMint(address to, uint256 tokenId, string memory uri) public onlyMember {
        _safeMint(to, tokenId);
        _setTokenURI(tokenId, uri);
        _totalSupply++; // Increment total supply upon minting
    }
    modifier onlyMember() {
        require(isMember[msg.sender], "Only members can call this function");
        _;
    }
    function totalSupply() external view returns (uint256) {
        return _totalSupply; // Return the total supply
    }
    // The following functions are overrides required by Solidity.
    function tokenURI(uint256 tokenId) public view override(ERC721, ERC721URIStorage) returns (string memory) {
        return super.tokenURI(tokenId);
    }
    function supportsInterface(bytes4 interfaceId) public view override(ERC721, ERC721URIStorage, IERC165) returns (bool) {
        return super.supportsInterface(interfaceId);
    }
    function addMember(address newMember) external onlyMember {
        require(!isMember[newMember], "Address is already a member");
        members.push(newMember);
        isMember[newMember] = true;
    }
```

b)

**Fig. 9** Governance tokens related smart contracts (a) Incentive logics contract (b) NFT contract.

### 5.2. Development of the Dapp Frontend

The front end of the DApp for the proposed system was developed using React JS, due to its flexibility, modular structure, and seamless compatibility with web3 JS, which facilitates interaction between the Ethereum blockchain and the web application. MetaMask integration enables users to connect their Ethereum wallets for blockchain transactions. As shown in Fig. 10, the DApp interface features four primary navigation tabs: Governance, Treasury, Maintenance, and User. The Governance section allows DAO participants to submit and vote on proposals related to building maintenance and improvement initiatives. Users can review active proposals, participate in voting processes, and monitor the implementation status of approved proposals. The Treasury tab displays financial information including the DAO's governance token holdings and Ethereum cryptocurrency balance, along with mechanisms to propose and vote on token allocation for incentivizing maintenance activities. The Maintenance tab serves as the central reporting hub where building occupants can submit maintenance requests by uploading images of issues, providing descriptions, and specifying locations. The User tab provides personalized information including available governance tokens, earned cryptocurrencies, and NFT badges reflecting the user's contribution history.

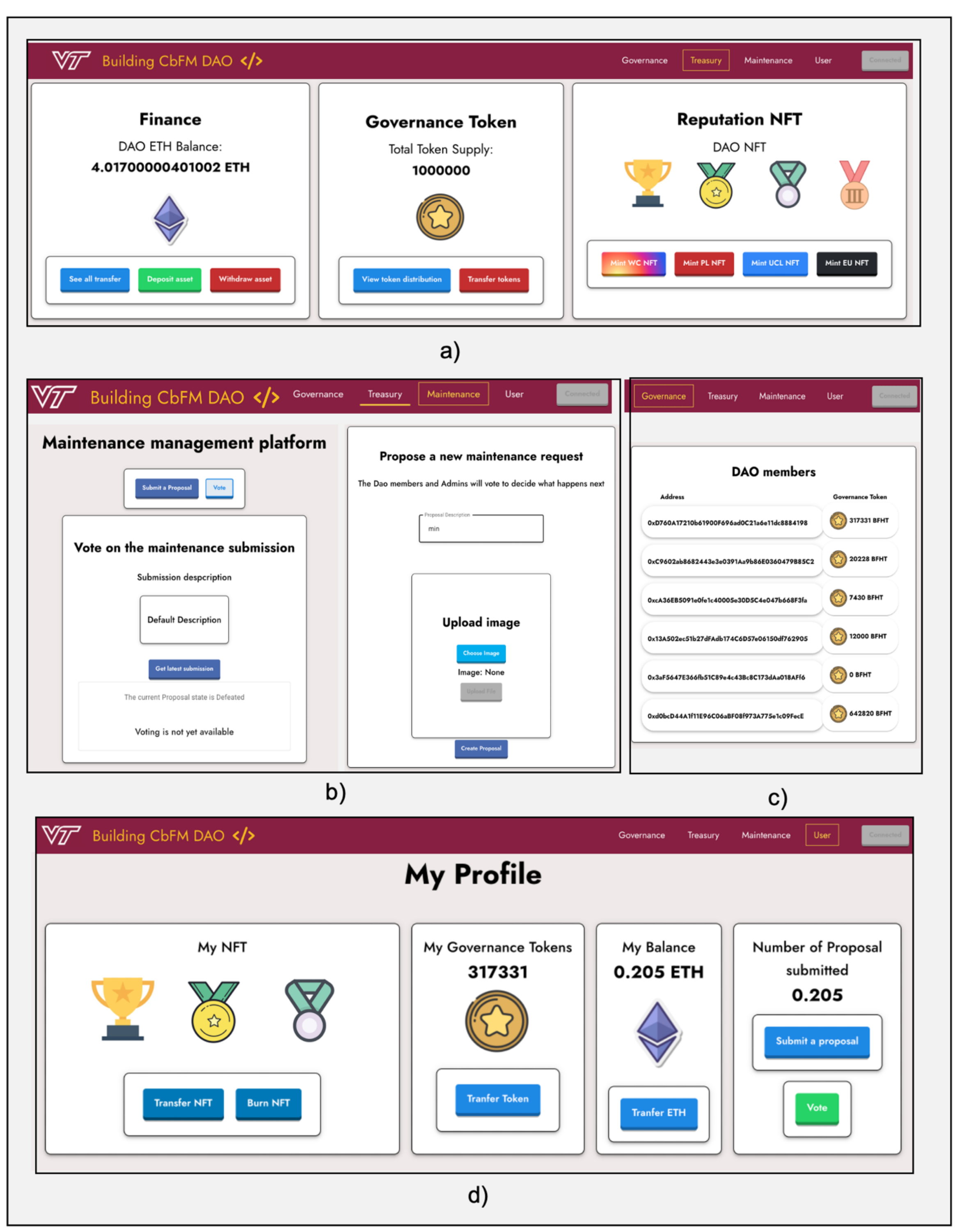

**Fig. 10** Frontend of the Maintenance management platform Dapp: (a) DAO Treasury tab (b) Maintenance management tab (c) Governance tab (d) User tab

## 6. Evaluation

This section outlines the evaluation and validation methodology used to assess the feasibility, usability, and usefulness of the proposed framework. A scenario-based evaluation approach was employed to simulate user interactions with the system. This validation method has been widely used in different blockchain-related studies (Tao et al. 2021, 2023; Naderi et al. 2023) and provides a feasible, effective method to demonstrate the viability of the technology in different practical contexts. The validation process was structured around several key scenarios, including user engagement with the maintenance management platform, and participation in the DAO governance through proposal creation and voting. Additionally, this study will evaluate the usability aspect of the proposed system using the System Usability Scale (SUS). The proposed system will also undergo qualitative assessment through expert interviews with researchers and facility managers to evaluate the platform's practical benefits and challenges for facility management applications.

### 6.1. Implementation setup

For implementation purposes, a Dapp was developed featuring three distinct stakeholders, each possessing an Ethereum account funded with 1 Sepolia testnet token. One participant deployed the DAO smart contract, including the DAO governor, governance, token, and Timelock contracts. A governance token designated as "BFHTokens" was created with a total supply of 1,000,000 units. Three accounts were each allocated 10,000 tokens, establishing their DAO membership status. Within this arrangement, one member was designated to initiate proposals, all three participated in the voting process, and one was responsible for proposal execution. In addition, user interactions with the Dapp followed the following process (Fig. 11). Steps (a)–(d) covered role assignments, account funding, and contract deployment. Steps (d) involved getting visual data from the submission. Steps (f)–(h) focused on DAO governance and incentivization.

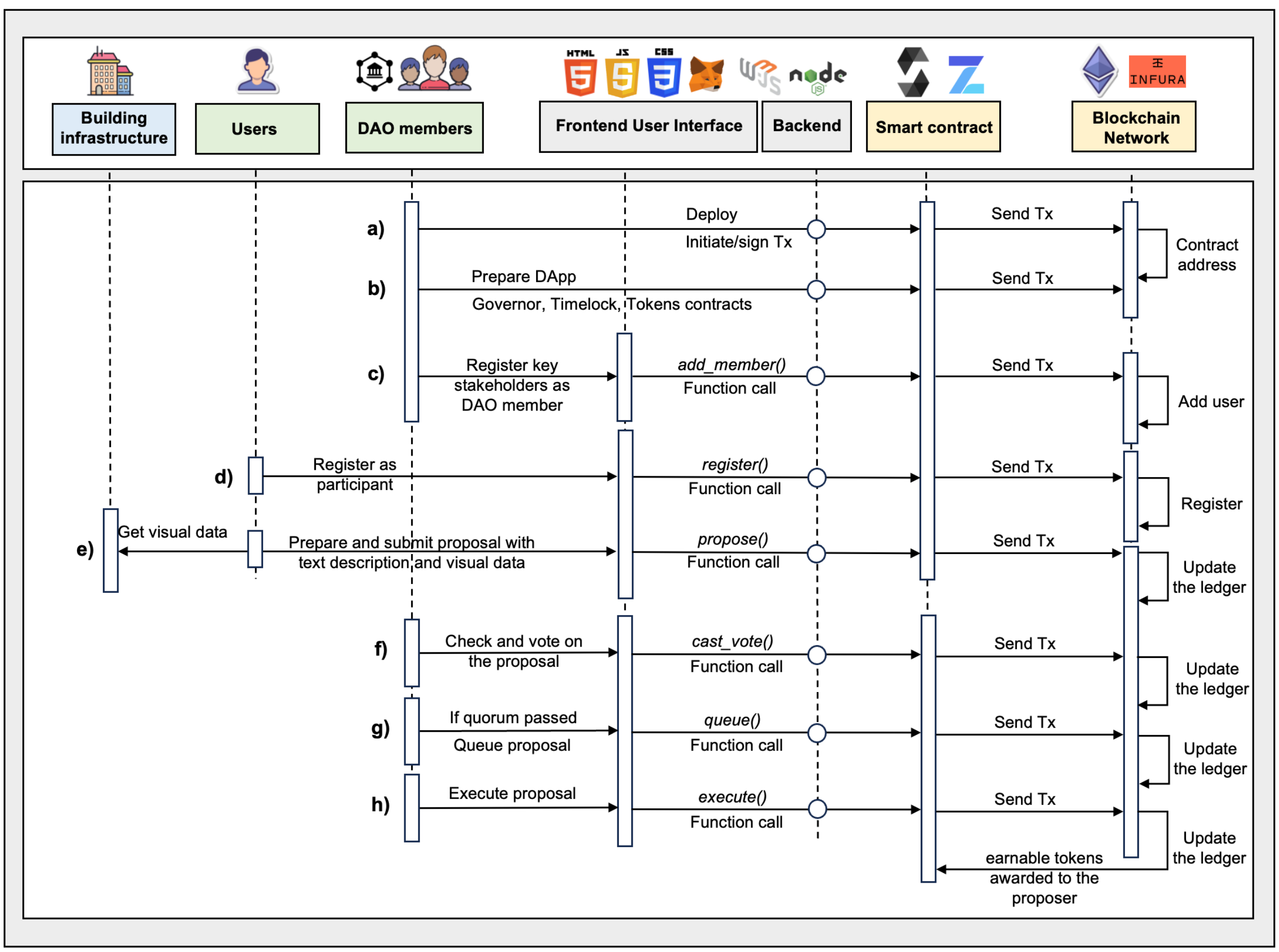

**Fig. 11** Sequence diagram of implementing the Dapp

### 6.2. Experiment 1: Proposal submission and incentivization workflow

This experiment aimed to evaluate the usage and workflow of the DAO-based maintenance management platform and governance process, from issue reporting through voting to incentivization. As illustrated in Fig. 12, the experiment follows a complete cycle of the decentralized facility management process. In the first step, a building occupant submits a maintenance request through the maintenance tab. The user describes the problem and uploads an image showing the damaged door. After completing the submission form, the user creates the proposal, which triggers a blockchain transaction that must be signed using their crypto wallet to confirm their identity and record the submission immutably on the blockchain. Once submitted, the building manager or DAO administrators can then access the submission through the voting interface. This interface presents the maintenance submission with its description and the uploaded image evidence, allowing decision-makers to assess the validity and priority of the reported issue. The DAO members can vote in favor, against, or abstain on the proposal. In this experiment, we simulated an approval scenario where the majority voted "In Favor" for addressing the reported door issue. The final step of the workflow demonstrates the incentivization mechanism, where the original proposer (the occupant who reported the broken door) receives a reward in the form of governance tokens.

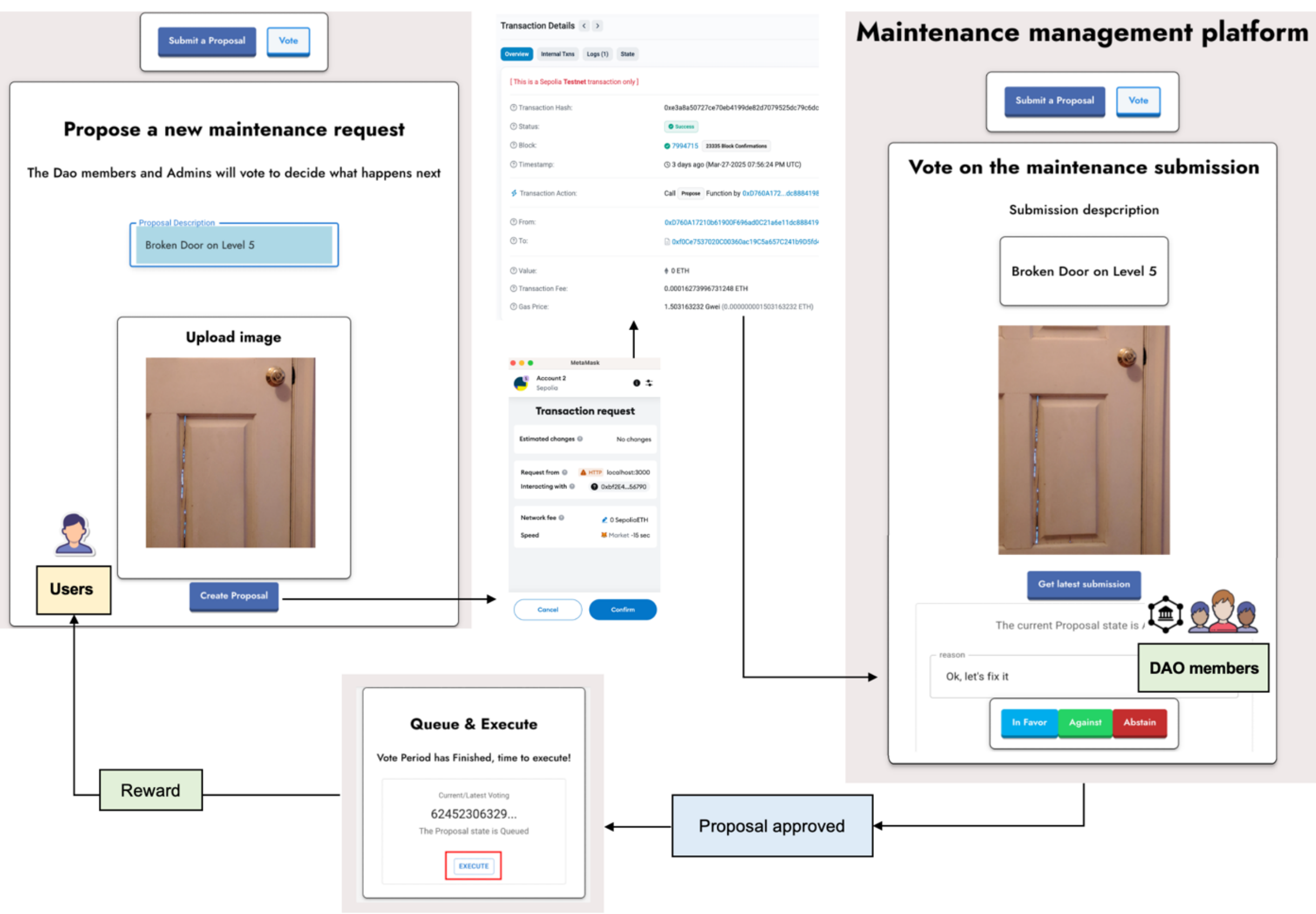

**Fig. 12** Proposal submission and incentivization workflow

### 6.3. Experiment 2: Modification of Token Incentive Distribution Mechanism.

This experiment aims to evaluate the system's decentralized governance infrastructure by testing the DAO's capacity to adjust the economic parameters governing participation rewards, specifically altering the token distribution algorithm that incentivizes community engagement and consensus-building activities. As illustrated in Fig. 17 (steps 1 and 2), a governance participant submitted a proposal to restructure the reward allocation formula, increasing the token incentive for governance participation to 500 tokens. Following submission, DAO stakeholders initiated the voting procedure (step 3), where members analyzed the potential economic impacts and cast weighted votes proportional to their governance token holdings. Upon achieving supermajority consensus through the voting mechanism, authorized token holders proceeded to queue and execute the approved modification, which updated the incentive parameters within the blockchain's governance contract (step 4). The revised token distribution metrics were subsequently verified as accurately implemented on-chain, as demonstrated in step 5.

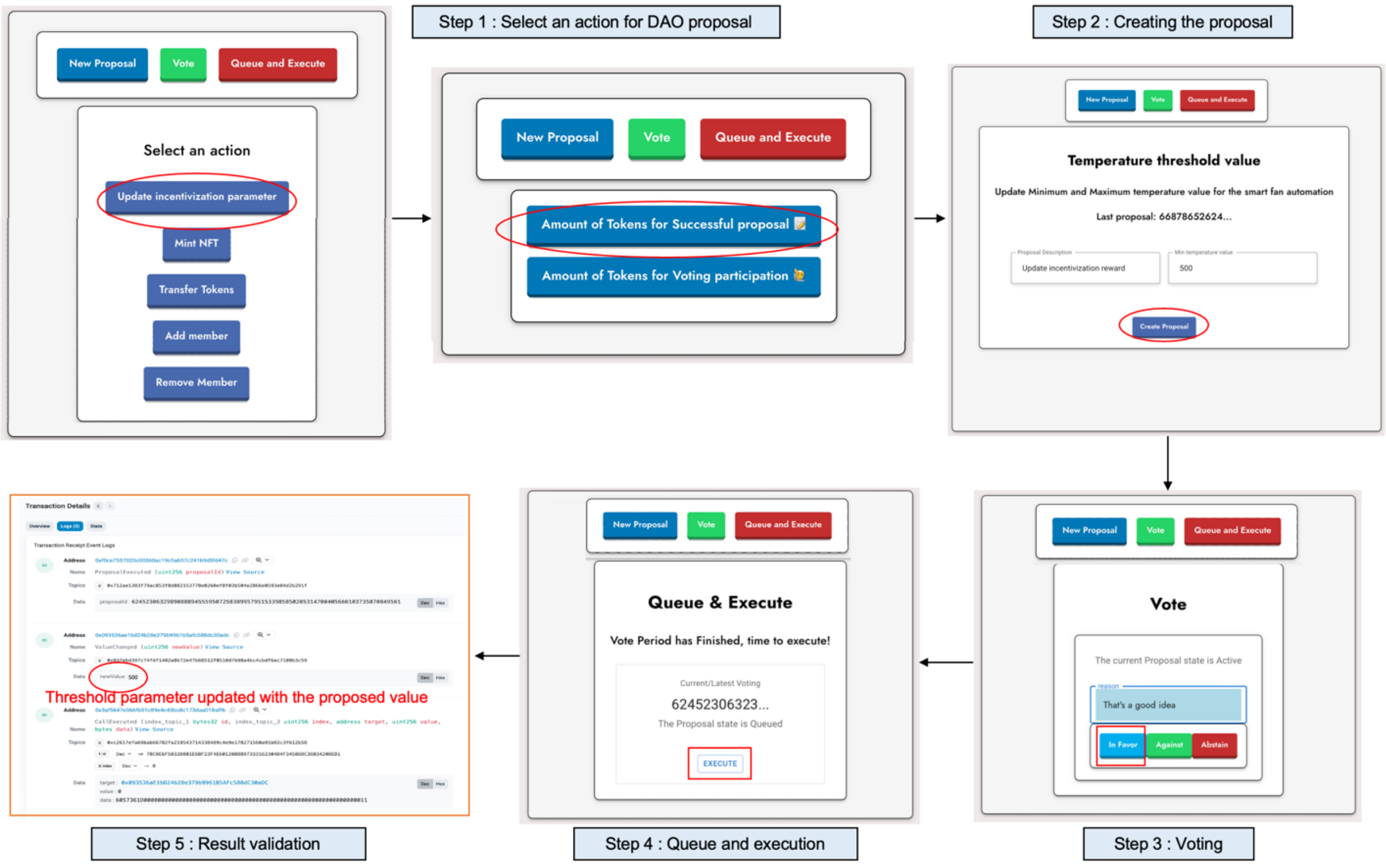

**Fig. 13** Modification of Token Incentive Setting workflow

### 6.4. System Usability evaluation

This study employed the System Usability Scale (SUS) to quantitatively assess the user-friendliness of the proposed system's key components: the decentralized governance platform and maintenance management platform. The testing involved 12 participants, which is higher than methodological approaches comparable to similar blockchain and DAO application studies (Dana et al. 2023; Peelam et al. 2025; Jeoung et al.) which typically utilized 10 participants for usability evaluation. As shown in Fig. 14. a) participants interacted with the platform by: (1) Proposing and voting on community maintenance issues and improvement suggestions. (2) Reviewing submitted issues with supporting documentation. (3) Engaging with the incentivization framework, including the distribution of tokens and NFTs for participation. Following these interactions, participants completed a post-experiment survey containing SUS statements (Appendix A and B).

### 6.5. Semi-structured interview

Semi-structured interviews were conducted with domain experts to gather comprehensive feedback on the decentralized governance and maintenance management platform. This qualitative assessment provided insights into the platform's benefits and challenges regarding usability, decision-making transparency, and effectiveness of the proposed blockchain incentives system. Five participants took part in the study. The participant count in this study is higher than in previous DAO governance research (Caviezel et al. 2023; Schmitten et al. 2023; Santos and Thesis), where typically 2-3 participants were consulted. The interviewees included two facilities managers and three researchers from Virginia Tech. As depicted in Fig. 14.b), the interviews were conducted via Zoom, with each session lasting approximately one hour. Interview recordings were automatically transcribed by Zoom for qualitative analysis. Sample interview questions are provided in Appendix C.

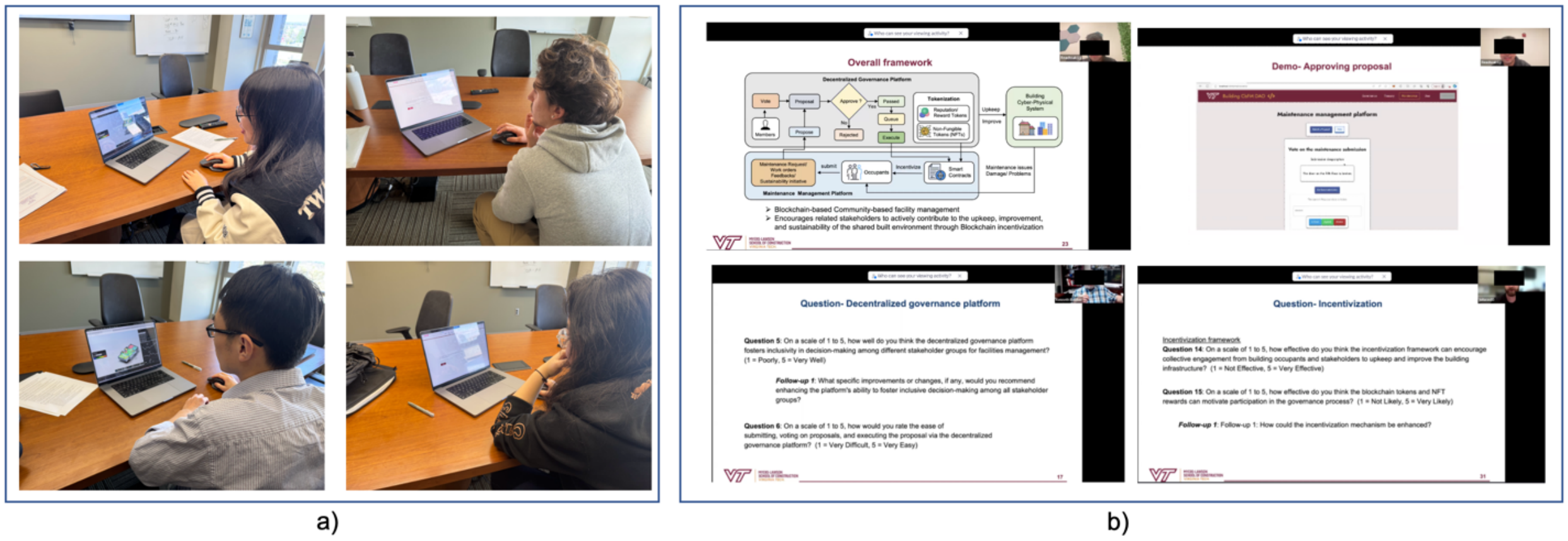

**Fig. 14** System Evaluation: (a) Usability Assessment (b) Semi-Structured Interview

## 7. Result and discussion

### 7.1. Cost analysis

The implementation of our blockchain-based incentivization system requires transaction fees on the Ethereum network. These fees, known as gas fees, cover the computational resources needed to process operations on the blockchain. Our cost evaluation revealed that deploying the core smart contracts (Governor, Timelock, and Token contracts) required approximately 0.051903 ETH, which translates to about USD 93.97. Regular transactions such as registering new building occupants, transferring incentive tokens, submitting maintenance proposals, voting on issues, and executing approved work orders range from $0.26 to USD 2.45 per transaction. The most frequent transaction—submitting maintenance reports—costs approximately USD 0.86. These costs were calculated during our testing phase on the Sepolia test network, providing a realistic estimate of expenses in a production environment. These fee calculations were performed during Sepolia testnet evaluations at an ETH rate of USD 1,810.47 (as of April 4th, 2025) and are itemized in the rightmost column of Table 3.

**Table 2** The transaction cost of the proposed decentralized governance platform

| Operations | Smart contract | Gas | Transaction fee (ETH) | Transaction fee (USD) |
|---|---|---|---|---|
| Contract deployment | DAO Governor | 3,880,388 | 0.003880 | 7.02 |
| Contract deployment | Timelock controller | 1,909,795 | 0.001909 | 3.46 |
| Contract deployment | GovernanceToken | 1,971,098 | 0.001971 | 3.57 |
| Contract deployment | NFTcontract | 1,505,175 | 0.001913 | 3.46 |
| Contract deployment | IncentiveLogic contract | 1,271,018 | 0.002921 | 5.29 |
| Adding DAO member | DAO Governor | 73,610 | 0.000110 | 0.20 |
| Proposal submission | DAO Governor | 108,168 | 0.000199 | 0.36 |
| Voting on proposal | DAO Governor | 93,186 | 0.000169 | 0.31 |
| Queuing proposal | DAO Governor | 123,769 | 0.000235 | 0.43 |
| Executing the Proposal | DAO Governor | 132,563 | 0.000238 | 0.43 |
| Governance Tokens transfer | GovernanceToken | 72,954 | 0.000139 | 0.25 |
| Ethereum tokens transfer | Timelock controller | 21,055 | 0.001052 | 1.90 |

### 7.2. Scalability

The system scalability is influenced by the limitations of the Ethereum blockchain, specifically its proof-of-stake consensus mechanism, which can limit the transaction throughput. Every transaction within the Ethereum network must receive validation from all participating nodes before being added to the blockchain. However, with the expansion of the network, the processes required to reach a consensus also escalate, potentially resulting in delays and increased gas fees. This challenge is common within the Ethereum blockchain-based systems, where throughput is limited to around 30 transactions per second (Abrol 2022). However, in this study's experimental setup, the decentralized governance helps distribute actions over time. For instance, it is quite improbable that all DAO members will simultaneously submit proposals, vote, or execute actions, which in turn reduce the likelihood of bottlenecks. However, if the system were to be adopted in a real-world scenario with a larger number of users, migrating to a more scalable blockchain solution like Polygon could be a practical solution.

### 7.3. Data Security and Privacy

This blockchain-based facility management platform leverages Ethereum's cryptographic foundation to create a balance between transparency and privacy. Rather than using personal identification, the system assigns users pseudonymous public keys, allowing them to participate in governance activities without exposing their identities (Wang et al. 2019b) . This means when occupants submit maintenance reports, vote on proposals, or receive incentive tokens, these activities are linked to their cryptographic signatures instead of personal information. Transaction validation in the system requires digital signatures with private keys, ensuring that only authorized individuals can interact with the platform. The platform intentionally makes certain information publicly viewable to foster community trust. Maintenance proposals, voting results, and token distribution records remain accessible to all participants, creating an environment where actions can be verified by anyone in the community. By combining pseudonymous identification with transparent processes, the system creates an environment where privacy concerns are addressed while maintaining the accountability necessary for effective community-based facility management.

### 7.4. Usability evaluation

This section presents the results from the usability testing of the proposed system with 12 participants, as shown in Fig. 15 and Table 3. The maintenance management platform received the highest average SUS score of 84.4, which according to the SUS interpretation framework by Bangor et al. (2009), falls within the "Excellent" adjective rating and corresponds to a "B+" grade on the SUS grading scale. This places the maintenance management platform well within the "Acceptable" range of usability, indicating that participants found this component highly intuitive and user-friendly.

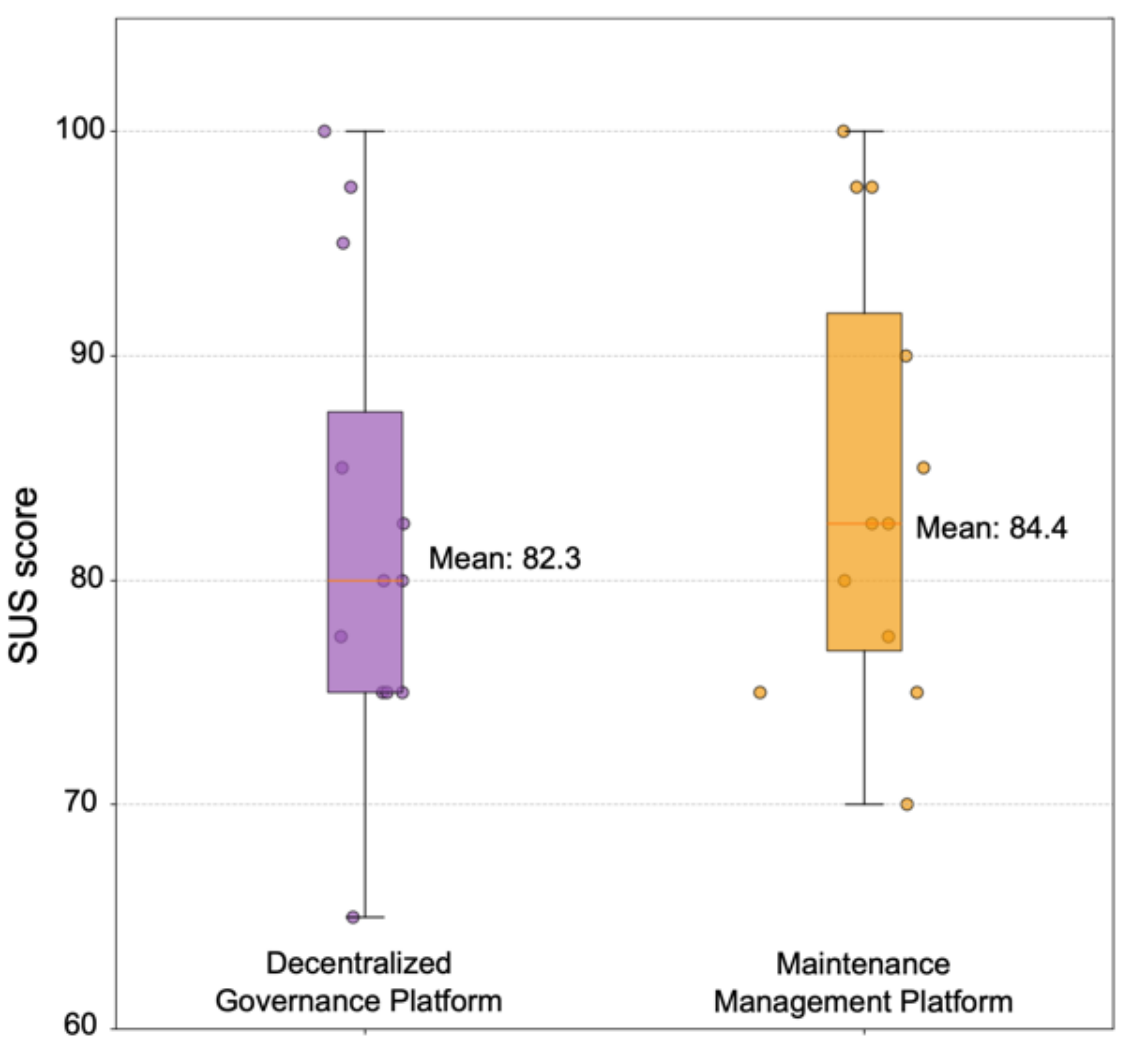

**Fig. 15** SUS score of the proposed system

The scores for the maintenance management platform ranged from 70 to 100, with the majority of participants rating the system between 75 and 97.5. This relatively consistent scoring suggests that users broadly agreed on the platform's usability, with few outliers in their assessments. The high usability rating aligns with the expert evaluation findings, where navigation/usability received a 4.5/5 rating from domain experts. The decentralized governance platform received an average SUS score of 82.3, which also falls within the "Excellent" adjective rating and corresponds to a "B" grade. The governance platform scores ranged from 65 to 100, with slightly more variability than the maintenance platform. This wider distribution may reflect the more complex nature of blockchain-based voting and governance mechanisms, which could present a steeper learning curve for some users. Despite this variability, the overall high score indicates that participants found the governance interface accessible and usable.

It's worth noting that both components achieved impressively high usability scores despite incorporating relatively complex blockchain technology. This suggests that the user interface design successfully abstracted the underlying technical complexity, allowing users to interact with the blockchain-based incentivization system without requiring deep technical knowledge.

**Table 3** System Usability Scale (SUS) questionnaire response

| SUS Statement (1=Strongly disagree, 2= Disagree, 3= Neutral, 4=Agree, 5 Strongly agree) | Maintenance Management Platform | | Gov. Platform | |
|---|---|---|---|---|
| | Mean | SD | Mean | SD |
| 1. I think that I would like to use this system frequently. | 4.50 | 0.5 | 4.41 | 0.49 |
| 2. I found the system unnecessarily complex. | 2 | 1 | 1.83 | 0.90 |
| 3. I thought the system was easy to use. | 4.58 | 0.49 | 4.42 | 0.64 |
| 4. I think that I would need the support of a technical person to use this. | 1.83 | 0.89 | 2 | 0.82 |
| 5. I found the various functions in this system were well integrated. | 4.67 | 0.47 | 4.50 | 0.5 |
| 6. I thought there was too much inconsistency in this system. | 1.66 | 0.47 | 1.83 | 0.55 |
| 7. I would imagine that most people would learn to use this system quickly. | 4.50 | 0.5 | 4.33 | 0.47 |
| 8. I found the system very cumbersome to use. | 1.66 | 0.47 | 1.75 | 0.43 |
| 9. I felt very confident using the system. | 4.67 | 0.47 | 4.41 | 0.49 |
| 10. I needed to learn a lot of things before I could get going with this system. | 2 | 0.57 | 1.75 | 0.59 |

### 7.5. Findings from semi-structured interview

The participants' insights on perceived benefits and challenges for each system component are illustrated in Fig. 16. Regarding the key benefits of the maintenance management platform and decentralized governance platform, all interviewees recognized enhanced building monitoring as the most significant advantage. This consensus highlights the fundamental value of enabling building occupants to help identify maintenance problems in large spaces, effectively distributing the responsibility of facility oversight.

Three additional benefits received strong recognition among the participants. For instance, the encouragement of collective upkeeping was viewed as a significant advantage, as respondents noted how the system effectively motivates building occupants to take ownership of their shared environment, even for issues that may not directly affect them personally. Also, establishing an NFT reputation system was highlighted as an innovative benefit. The NFT reputation badges were particularly valued for their potential to serve as verifiable credentials that could follow users across different contexts. Enabling preventative maintenance was recognized as a practical benefit with significant operational implications. The ability to identify maintenance issues before they escalate to regulatory violations or major failures was particularly valued by facility management professionals. The blockchain-based immutable record keeping was appreciated for its transparency and resistance to tampering, which addresses traditional challenges in maintenance documentation. Behavioral improvement through incentives was recognized as an effective mechanism for engaging occupants who might otherwise ignore issues that don't directly affect them. Cost and resource efficiency, while mentioned less frequently, was still acknowledged as a valuable outcome of distributed monitoring.

The evaluation also revealed several challenges and limitations. Token utility and real-world value exchange emerged as the most significant concerns, with participants questioning how blockchain tokens would translate to tangible benefits for users. This suggests that clear value propositions and exchange mechanisms must be established for successful implementation. Technical understanding gaps and voting power distribution concerns also represent significant barriers to adoption. This indicates the need for educational components for users and careful governance design. Two interviewees mentioned the need for an improved feedback loop, suggesting improvements in communicating how user reports lead to actual facility improvements. Only one participant explicitly mentioned blockchain transaction delays.

Quantitative ratings further illuminate the system's strengths and limitations. Navigation/usability received an impressive 4.5/5 rating, suggesting the interface design successfully achieves its goals despite the technical complexity of blockchain technology. Inclusiveness also scored highly (4.5/5), indicating the system effectively enables broad participation in facility management decisions. The effectiveness for building upkeep (3.8/5) and blockchain/NFT effectiveness for incentivization (3.7/5) both received moderately strong ratings, suggesting conditional success dependent on implementation context. The implementation likelihood/adoption potential rating (3.7/5) reveals cautious optimism about real-world deployment, with academic interviewees generally more optimistic than facility management practitioners. These results suggest that while the blockchain-based facility management system shows considerable promise, particularly in interface design and inclusive participation, successful implementation will require addressing token utility concerns and bridging the gap between theoretical potential and practical facility management requirements.

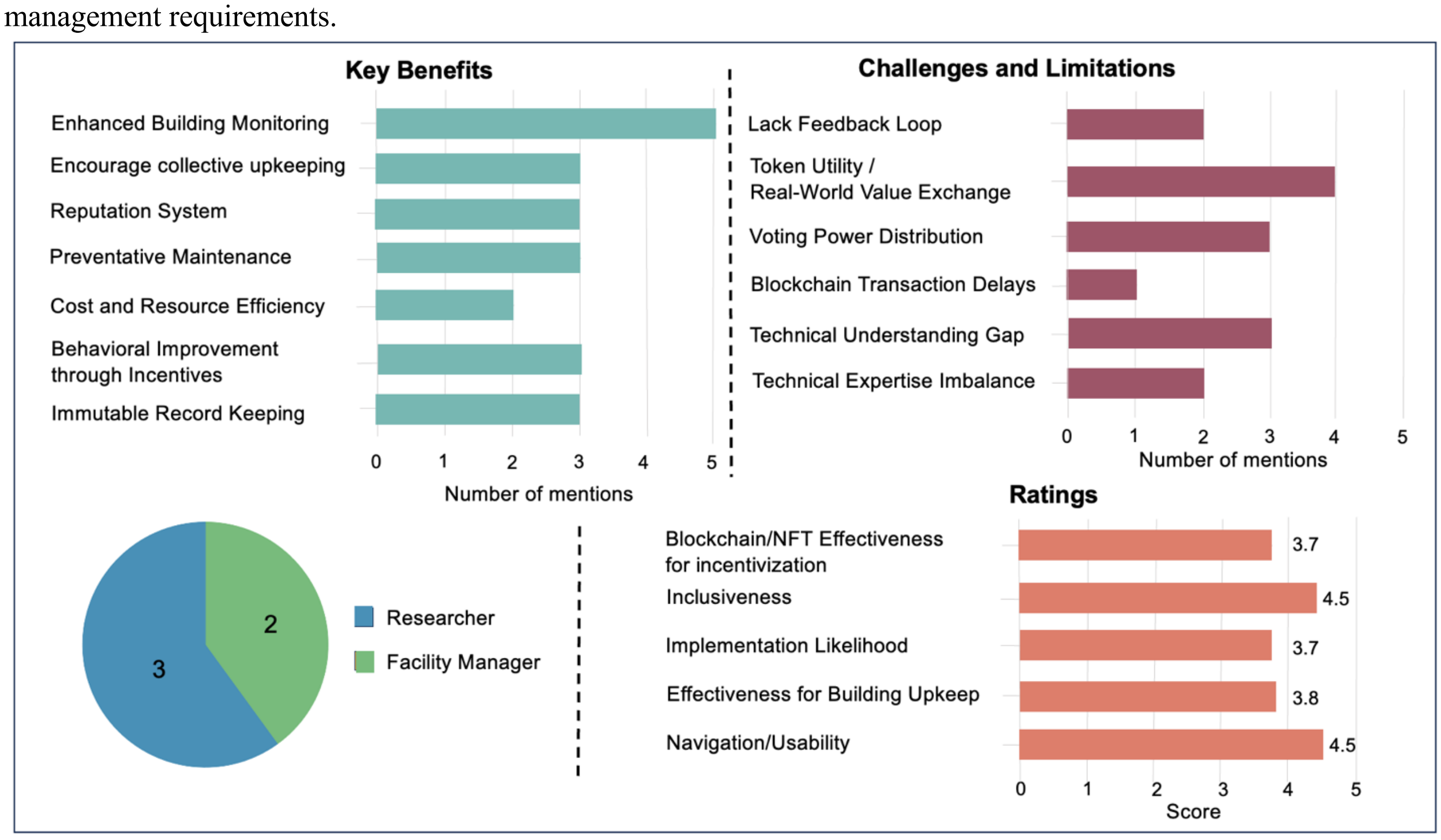

**Fig. 16** Findings from the semi-structured interview

### 7.6. Novelty and originality

Figure 17 illustrates how each feature of the proposed system directly addresses a corresponding challenge in current practices mentioned in section 2.2. For instance, the decentralization feature of the proposed system eliminates the centralization problem in the current CbFM practice by distributing decision-making power across a peer-to-peer network instead of concentrating it on a single authority. The blockchain-based voting mechanism within the decentralized governance platform also tackles the lack of inclusiveness by enabling democratic, community-driven decision-making. In addition, the issue of enforceability is tackled through smart contracts, which automate processes and ensure compliance, overcoming the current lack of trust and weak accountability. For instance, the automation and autonomy of task execution, such as the release of incentives to a particular party when the condition is met. Also, the transparency in the incentivization and decision-making process is significantly improved in the proposed system as it provides data traceability and immutability. To address insufficient motivation and lack of antifraud measures, our proposed system implements fungible and non-fungible token-based (FT and NFT) rewards to incentivize active participation. Lastly, the blockchain's immutable and auditable record-keeping capabilities solve the potential for manipulation or loss of records in current systems, ensuring a trustworthy and transparent management process. These improvements collectively represent a substantial advancement over the state-of-the-art in community-based facilities management.

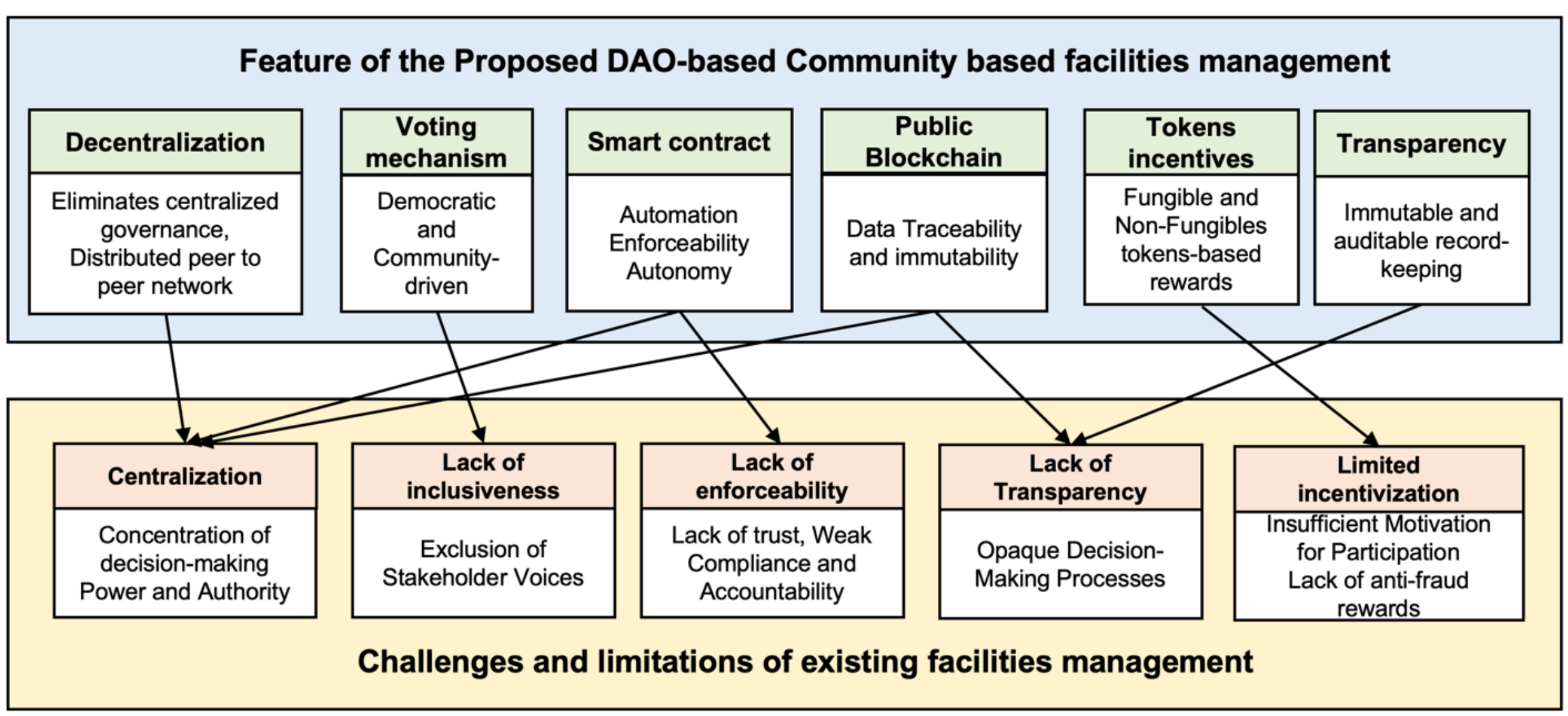

**Fig. 17** Challenges and limitations of existing facilities management practices with the features of the proposed DAO-based community facilities management system

### 7.7. Limitations and Future Research

This section addresses the key constraints of the proposed system and outlines potential directions for future research. A significant challenge in the current implementation stems from its dependence on Ethereum cryptocurrency for system transactions. The characteristic price volatility of Ethereum creates financial unpredictability, complicating expense management for both users and DAO members. This inconsistency between projected and actual costs could impede widespread adoption. Future system iterations might resolve this issue through the integration of stablecoins like USDT or USDC (Thanh et al. 2023), which provide more consistent decentralized payment options by maintaining value equivalence with reserve assets such as the U.S. Dollar. A notable constraint of the current research is its validation through simulated case studies rather than real-world implementations. Future work should investigate practical deployments in actual building environments to evaluate performance under authentic conditions. Future research could also explore AI-driven automation for blockchain-based governance tasks. Rather than requiring manual execution of blockchain operations by DAO members, an AI assistant could be developed to streamline these processes, allowing users to manage smart contract executions and governance decisions through simple voice or text commands.

## 8. Conclusion

This paper presents a novel blockchain and DAO's decentralized governance and incentivization framework for smart building community-based facilities management. The proposed framework comprises several key components. The decentralized governance platform, powered by DAO's governance, facilitates transparent decision-making and resource management. The maintenance management platform contains the incentivization framework that encourages occupants to report any building maintenance issues and/or provide relevant feedback in contributing to the maintenance and enhancement of shared building infrastructure. The resource and code implementation for these components is available on a GitHub repository under an open-source license (Ly 2024), allowing for further development and application of this framework beyond autonomous building management.

This study contributes to the body of knowledge in several ways: (1) Providing a novel DAO-based decentralized governance model tailored for facility management, empowering stakeholders with collective decision-making capabilities. (2) Introducing an incentivized framework for community-based facility management, encouraging active participation and contributions from building occupants, tenants, and other stakeholders. (3) Developing a full-stack, open-source DApp that serves as a template for other blockchain-related applications in the domain of decentralized facility management and maintenance. (4) Offering insights into the perceived level of inclusiveness and decentralization achieved through the proposed system, based on qualitative feedback obtained from user studies. (5) Demonstrating the practical implementation and evaluation of a DAO and blockchain-based system in a real-world smart building environment, contributing to the understanding of the challenges and opportunities associated with such solutions in the built environment.

The evaluations of the system included analyses of cost efficiency, scalability of the governance and incentivization system, data security, and privacy. This study also evaluates the system's usability System Usability Scale (SUS). Semi-structured interviews with researchers and facility managers were also conducted to evaluate the platform's practical benefits and challenges. The results from these evaluations demonstrated that the developed prototype system can potentially serve as the viable framework for future incentivization systems for community-based facilities management in building infrastructure.

**Acknowledgments**
This research has received no external funding.

**Availability of data and materials**
Data will be made available on request.

**Competing interests**
The authors have no conflicts of interest to declare that are relevant to the content of this article.

**Funding**
No funding was received to assist with the preparation of this manuscript.

**Authors' contributions**
Supervision: [Alireza Shojaei]; Conceptualization: [Reachsak Ly]; Writing - original draft preparation: [Reachsak Ly]; Writing - review and editing: [Reachsak Ly]; Visualization: [Reachsak Ly]; Methodology: [Alireza Shojaei]; Project administration: [Alireza Shojaei]

**Appendix A System Usability Scale (SUS) Questionnaire for the Maintenance Management System of the Blockchain-based incentivization platform for Community-Based Facilities Management within smart buildings.**

| Modified SUS Statement for user experience evaluation of the Maintenance Management System | (1=Strongly disagree, 2= Disagree, 3= Neutral, 4=Agree, 5 Strongly agree) | | | | |
|---|---|---|---|---|---|
| | 1 | 2 | 3 | 4 | 5 |
| 1. I think that I would like to use this Maintenance Management System frequently for submitting maintenance request in smart building. | | | | | |
| 2. I found the Maintenance Management System unnecessarily complex. | | | | | |
| 3. I thought the Maintenance Management System was easy to use. | | | | | |
| 4. I think that I would need the support of a technical person to use this Maintenance Management System. | | | | | |
| 5. I found the various functions in this Maintenance Management System were well integrated. | | | | | |
| 6. I thought that there was too much inconsistency in this Maintenance Management System. | | | | | |
| 7. I imagine that most people would learn to use this Maintenance Management System very quickly. | | | | | |
| 8. I found the Maintenance Management System very awkward to use. | | | | | |
| 9. I felt very confident using the Maintenance Management System. | | | | | |
| 10. I needed to learn a lot of things before I could get going with this Maintenance Management System. | | | | | |

**Appendix B System Usability Scale (SUS) Questionnaire for the Decentralized governance platform of the Blockchain-based incentivization platform for Community-Based Facilities Management within smart buildings.**

| Modified SUS Statement for user experience evaluation of the Decentralized governance platform | (1=Strongly disagree, 2= Disagree, 3= Neutral, 4=Agree, 5 Strongly agree) | | | | |
|---|---|---|---|---|---|
| | 1 | 2 | 3 | 4 | 5 |
| 1. I think that I would like to use this Decentralized governance platform frequently for providing incentives to the submitted maintenances request. | | | | | |
| 2. I found the Decentralized governance platform unnecessarily complex. | | | | | |
| 3. I thought the Decentralized governance platform was easy to use. | | | | | |
| 4. I think that I would need the support of a technical person to use this Decentralized governance platform. | | | | | |
| 5. I found the various functions in this Decentralized governance platform were well integrated. | | | | | |
| 6. I thought that there was too much inconsistency in this Decentralized governance platform. | | | | | |
| 7. I imagine that most people would learn to use this Decentralized governance platform very quickly. | | | | | |
| 8. I found the Decentralized governance platform very awkward to use. | | | | | |
| 9. I felt very confident using the Decentralized governance platform. | | | | | |
| 10. I needed to learn a lot of things before I could get going with this Decentralized governance platform. | | | | | |

**Appendix C Interview questions for the blockchain-based incentivization and decentralized governance platform for community-based facilities management**

Theme 1: Usability of the platform

- Question 11: On a scale of 1 to 5, how easy is it to navigate the maintenance management platform for submitting the maintenance request? (1 = Very Difficult, 5 = Very Easy)
- Question 12: On a scale of 1 to 5, how would you rate the ease of voting on proposals and executing the proposal for the incentivization within the decentralized governance platform? (1 = Very Difficult, 5 = Very Easy)
- Question 13: On a scale of 1 to 5, how intuitive do you find the overall platform's interface? (1 = Not Intuitive, 5 = Very Intuitive)
    - Follow-up 1: What specific aspects make it intuitive or non-intuitive?
    - Follow-up 2: What changes should be made, if any, to improve the usability?

Theme 2: Incentivization framework

- Question 14: On a scale of 1 to 5, how effective do you think the incentivization framework can encourage collective engagement from building occupants and stakeholders to upkeep and improve the building infrastructure? (1 = Not Effective, 5 = Very Effective)
- Question 15: On a scale of 1 to 5, how effective do you think the blockchain tokens and NFT rewards can motivate participation in the governance process? (1 = Not Likely, 5 = Very Likely)
    - Follow-up 1: How could the incentivization mechanism be enhanced?

Theme 3: Inclusivity in decision-making of the blockchain-based incentivization decentralized governance platform.

- Question 16: On a scale of 1 to 5, how well do you think the decentralized governance platform fosters inclusivity in decision-making among different stakeholder groups for the incentivization process? (1 = Poorly, 5 = Very Well)
    - Follow-up 1: What specific improvements or modifications to the platform, if any, would you recommend to foster inclusive decision-making for the incentivization process?

Theme 4: Benefits and challenges

- Question 17: What do you see as the main benefits of using this incentivization and governance framework for encouraging collective upkeeping of building infrastructure?
- Question 18: What are the key challenges or limitations that you foresee in implementing this platform?

- Follow-up 1: Overall, what specific improvements or modifications to the system would you recommend, if any, to enhance the collective participation of building stakeholders in the upkeeping and improvement of building infrastructure?

Theme 5: Adoption potential

- Question 19: On a scale of 1 to 5, Please rate the following aspects of the platform (1 = Very Low, 5 = Very High)
    - The likelihood of implementing this system for future building infrastructure.
    - The platform's effectiveness in fostering collective upkeeping and improvement of building infrastructure.